\documentclass[reprint, superscriptaddress, aps, showkeys, pra, twocolumn]{revtex4-2}
\usepackage{graphicx} 
\usepackage{amsmath, amssymb}
\usepackage{mathtools}
\usepackage{easybmat}
\usepackage{braket}
\usepackage[colorlinks=true,citecolor=blue,urlcolor=blue, linkcolor=magenta]{hyperref}

\newcommand{\pt}{\mathcal{PT-}}

\begin{document}

\title{$\mathcal{PT-}$Symmetric Two-Level Open Quantum Systems: Information Theoretic Facets}
\author{Baibhab Bose\textsuperscript{}}
\email{Corresponding author; baibhab.1@iitj.ac.in}
\author{Devvrat Tiwari\textsuperscript{}}
\email{devvrat.1@iitj.ac.in}
\author{Subhashish Banerjee\textsuperscript{}}
\email{subhashish@iitj.ac.in}
\affiliation{Indian Institute of Technology Jodhpur-342030, India\textsuperscript{}}
\date{\today}

\begin{abstract}
    The theory of a two-level $\eta$-Hermitian Hamiltonian with $\mathcal{PT}$ symmetry is reviewed and extended to include open system dynamics. A first-principles derivation of the generalized Gorini-Kossakowski-Sudarshan-Lindblad master equation appropriate for a $\mathcal{PT}-$symmetric Hamiltonian is presented. Inspired by a simple light matter interaction open system model, information theoretic quantities like a non-Markovian witness and fidelity are calculated for the $\mathcal{PT-}$symmetric Hamiltonian, and the results are compared with their corresponding two-level Hermitian counterparts. The nature of entanglement between two $\mathcal{PT-}$symmetric and Hermitian open quantum systems is calculated, and the contrast observed.
\end{abstract}
\maketitle

\section{Introduction}
In usual quantum mechanics, Hamiltonians that are dealt with are Hermitian so as to keep the eigenvalues real. It was shown in~\cite{Bender1, Bender2_1999, Bender_3} that a non-Hermitian Hamiltonian with parity and time-reversal ($\mathcal{PT}$) symmetry also possesses a real spectrum in a specific regime of the values of the parameters of the Hamiltonian. This hints at the fact that $\mathcal{PT-}$symmetry is a more general symmetry for quantum mechanical Hamiltonians as compared to Hermiticity, and the key features of quantum mechanics should be re-evaluated in light of this $\mathcal{PT-}$symmetry to find out the physical significance of such systems. Subsequently, a number of advancements have been made in $\mathcal{PT-}$symmetric non-Hermitian theory~\cite{Bender_4_2003, Bender_5_2005, Bender_6_2007, Asok_das_2010, neutosci1_PhysRevD.89.125014, Nori_2019, Das_Bhasin_2025, Kumari_2022}. The necessary and sufficient conditions for non-Hermitian Hamiltonians to have real spectra were investigated in a number of works~\cite{Mostafazadeh_1_2002, Mostafazadeh_2_2002, Mostafazadeh_3_2002, Mostafazadeh_4_2004, Mostaf_5_doi:10.1142/S0219887810004816}. A linear positive-definite operator $\eta$ can be obtained from the parameters of the non-Hermitian Hamiltonian such that the Hermiticity is supplanted by a more general class of $\eta$-Hermiticity, $\eta^{-1}H^{\dagger}\eta=H$, maintaining the energy eigenvalues real.
The non-Hermitian Hamiltonians can also be rotated by another Hermitian matrix $G$ ($G^\dagger G=\eta$) to a Hermitian Hamiltonian. 

Numerous fields, including optics~\cite{Optics_Rter2010}, electronics~\cite{Electronics_PhysRevA.84.040101}, microwaves~\cite{microwaves_PhysRevLett.108.024101}, mechanics~\cite{mechanical_Bender2013}, acoustics~\cite{acoustic_Fleury2015}, atomic physics~\cite{atomic_1_Baker1984,atomic_2_PhysRevLett.117.123601,atomic_1_PhysRevLett.110.083604}, quantum optics~\cite{Javid_2019, Agarwal_2012} and single-spin systems~\cite{singlespin_doi:10.1126/science.aaw8205}, among others have found use for non-Hermitian Hamiltonians with $\mathcal{PT-}$symmetry. 
Further, in the fields of elementary particle physics and quantum field theories, $\mathcal{PT-}$symmetric Hamiltonians have found widespread application. The framework of non-Hermitian theories with $\mathcal{PT}$ symmetry has been used to study a number of other fundamentally significant issues, such as neutrino oscillations~\cite{neutosci_2_Ohlsson2016} neutrino mass generation~\cite{neutmassgen_Alexandre2015}, light neutrino masses in Yukawa theory~\cite{Yukawa_Alexandre2017}, spontaneous symmetry breaking and the Goldstone theorem~\cite{Goldstone_PhysRevD.98.045001}, and the Brout–Englert–Higgs mechanism~\cite{Brout_higgs_PhysRevD.99.045006, brout_higgs_2_PhysRevD.99.075024}. 

The theory of open quantum systems deals with the analysis of a quantum system interacting with a surrounding bath~\cite {Weiss2011, Breuer2007, Banerjee2018, Omkar2016, Vacchini_2011, Tiwari_2023, tiwari2024strong}. A well-known canonical open system approach is the Gorini-Kossakowski-Sudarshan-Lindblad (GKSL) type interaction modeling Markovian evolution~\cite{GKLSpaper, Lindblad1976}. Beyond the GKSL type interaction, non-Markovian evolution has been extensively studied in recent years~\cite{Hall_2014, Rivas_2014, RevModPhys.88.021002, CHRUSCINSKI20221, banerjeepetrucione, vega_alonso, Utagi2020, kading2025}. In the context of open quantum systems, various facets of quantum correlations~\cite{Chakrabarty2011, Javid_Dutta_2019, Javid_Alok_2018} and quantum walks~\cite{SB_2007, SB_2008, SB_2010}, an algorithm to simulate quantum dynamics, have been studied. $\mathcal{PT-}$symmetric systems in the context of quantum correlations~\cite{Naikoo_2021, Javid_SB_2019} and quantum walk~\cite{Badhani_2024} have been studied.

The GKSL master equation was adapted to a $\mathcal{PT-}$symmetric evolution~\cite{Scolarici2006, Scolarici2007, TommyOhlsson_densitymatrixforma_PhysRevA.103.022218, kleefeld_2009arXiv0906.1011K}. In this paper, we discuss thoroughly the $\eta$ inner product eigenspace and how the dual space is constructed out of the corresponding eigenvectors. This is distinct from the usual algebra of Hermitian quantum mechanics. The quantum mechanical observables and the density matrix are modified by introducing the $\eta$ metric, which depends on the parameters of the Hamiltonian. More specifically, the modification of the projectors of this $\mathcal{PT-}$symmetric eigenspace by $\eta$ is shown. This modifies the density matrix of the system to a generalized density matrix, which enables us to study the open system dynamics of a $\mathcal{PT}-$symmetric system. A GKSL master equation suited for this $\eta$ folded space is derived from first principles for Markovian evolution. After that, we proceed to a more general case of non-Markovian evolution using the modified CPTP map to numerically calculate the generalized density matrix. 
We review the two-level quantum system with a general $\mathcal{PT-}$symmetric non-Hermitian Hamiltonian. We model a scenario where this $\mathcal{PT-}$symmetric non-Hermitian Hamiltonian is the system of interest and interacts with a single-mode field by a Jaynes-Cummings type~\cite{Larson_Dicke} interaction. This allows us to envisage the $\mathcal{PT-}$symmetric non-Hermitian Hamiltonian system as an open quantum system. Further, the coupling strength is taken arbitrarily strong, making it conducive to observing non-Markovian evolution. In \cite{Frith_2020}, a $\mathcal{PT}-$symmetric Hamiltonian of a closed system of the Jaynes-Cummings model was dealt with. There, the interaction Hamiltonian between the spin and radiation mode was taken as imaginary to make the total system Hamiltonian $\mathcal{PT}-$symmetric. Here, we take a $\mathcal{PT}-$ symmetric Hamiltonian as our system and couple it with a Hermitian radiation bath, which also renders the interaction Hamiltonian $\mathcal{PT}-$symmetric.

For a Hermitian Hamiltonian of a two-level (spin) system or a qubit, various information-theoretic quantities have been used to gain a broad picture of the given dynamics. The Breuer-Laine-Piilo (BLP) measure is a well-known witness of the non-Markovianity of the evolution~\cite{lainebreuer}. The fidelity is another such quantity and is a measure of the overlap between two quantum states~\cite{Nielsen_Chuang_2010, Jozsa1994, Uhlmann1976}. Here, we take a two-level $\mathcal{PT-}$symmetric system and modify the above measures to accommodate the $\eta$-Hermiticity of the $\mathcal{PT-}$symmetric evolution and compare these quantities with those of a Hermitian two-level system.
Concurrence is a measure of the entanglement growth between two qubits~\cite{Nielsen_Chuang_2010, Wootters1998}. We redefine it for the $\mathcal{PT-}$symmetric case and also for an open system scenario of Tavis-Cummings type~\cite{Larson_Dicke} interaction with the bath.  We observe entanglement growth between two $\mathcal{PT-}$symmetric systems coupled to each other in a manner analogous to the interaction between two spins of an integrable system of Ising spin chain~\cite{Arul_2007}. We compare it with the concurrence of two Hermitian spin systems having a $\sigma^z-\sigma^z$ type interaction under the same open system setup to get a deeper look at how the $\eta$-Hermiticity and the parameters of the $\mathcal{PT-}$symmetric Hamiltonian affect entanglement. In ~\cite{Stenholm01042004}, the non-Hermitian aspect of the generator of the time evolution is shown in the context of time reversal in open quantum systems. However, in this work, our aim is to have a consistent framework for $\pt$symmetric open quantum systems, where the system is itself $\pt$symmetric and interacts with its ambient environment.

The paper is organized as follows. In Sec.~\ref{SecII}, we discuss the $\eta$-Hermiticity of a $\mathcal{PT-}$symmetric Hamiltonian. Section~\ref{SecIII} discusses the dynamics of an open system for a $\mathcal{PT-}$symmetric Hamiltonian. The information-theoretic study of a non-Hermitian open system and comparison with its Hermitian counterpart is done in Sec. \ref {SecIV}. The entanglement between two $\mathcal{PT-}$symmetric open systems is studied in Sec. \ref{SecV}, followed by conclusions in Sec. \ref{SecVI}.

\section{$\eta$- Hermiticity of a $\mathcal{PT-}$symmetric Hamiltonian}\label{SecII}
The general form of a non-Hermitian Hamiltonian which is $\mathcal{PT-}$symmetric is \cite{Bender2002,Bender_2004}
\begin{align}\label{Hohl}
    \mathcal{H} = \begin{pmatrix}
    re^{i\psi} & se^{i\phi} \\
    se^{-i\phi} & re^{-i\psi}
    \end{pmatrix}.
\end{align}
To guarantee the orthogonality of the eigenstate vectors (with respect to the $\mathcal{PT}$ or $\mathcal{CPT}$ inner products), one can demand $H$ to be symmetric, choosing $\phi=0$, i.e.,
\begin{align}{\label{nonhermiH}}
    H = \begin{pmatrix}
    re^{i\psi} & s \\
    s & re^{-i\psi}
    \end{pmatrix},
\end{align}
where $r$, $s$ and $\psi$ are real parameters. Such a Hamiltonian can be obtained from an effective two-level system with balanced gain and loss conditions~\cite{Du_2018}.
The $\mathcal{P}$ operator acting on $H$ yields
\begin{align}
    \mathcal{P}H\mathcal{P}^{-1}=\begin{pmatrix}
    re^{-i\psi} & s \\
    s & re^{i\psi}
    \end{pmatrix}=H^{\dagger}.
\end{align}
If the complex conjugation operator ($\mathcal{T}H\mathcal{T}^{-1}=H^*$) is applied, we get back $H$. This proves that $(\mathcal{PT})H(\mathcal{PT})^{-1}=H$.

The $\mathcal{PT}$ symmetric Hamiltonian is also $\eta$-Hermitian~\cite{Mostafazadeh_2002}. The corresponding vector space is defined using an inner product characterized by the $\eta$-metric, such that $\bra{~\cdot~}\eta\ket{~\cdot~}=1$ and is denoted as $\eta$-inner product space.
A $\mathcal{PT-}$symmetric Hamiltonian, e.g., $H$ in \eqref{nonhermiH}, is $\eta$-Hermitian for a specific $\eta$.
The Hamiltonian \eqref{nonhermiH} has eigenvalues ($E_n$),
\begin{align}
    E_1=r\cos{\psi}-\sqrt{s^2-r^2\sin^2{\psi}}, \nonumber \\
    E_2=r\cos{\psi}+\sqrt{s^2-r^2\sin^2{\psi}}.
\end{align}
For $s^2\geq r^2 \sin^2{\psi}$, these eigenvalues are real. 
The eigenvectors of a non-Hermitian Hamiltonian are different from those of a Hermitian Hamiltonian, as the former have different left and right eigenvectors. This means that the complex conjugation operation on an eigenvector doesn't yield its left eigenvector. 

We now briefly discuss the properties of an $\eta$-Hermitian Hamiltonian. The eigenvalue equation for a $\eta$-Hermitian Hamiltonian is
\begin{align}\label{eigenvalue equation}
    H\ket{E_n^R}&=E_n\ket{E_n^R}, \nonumber \\ 
    \bra{E_n^L}H&=E_n\bra{E_n^L}.
\end{align}
The left eigenvector for $H$ in \eqref{nonhermiH} is explained in Eq. \eqref{lefteigenvecs} in Appendix A.
Here $\ket{E^R}^{\dagger}=\bra{E_n^R}\neq\bra{E_n^L}$.
From the above eigenvector equations, it can be obtained that
\begin{align}\label{InnerP}
    \langle{E_m^L}\ket{E_n^R}&=\delta_{mn}, \nonumber \\
    \bra{E_m^R}E_n^L\rangle&= \delta_{mn},
\end{align}
as shown in ~\cite{kleefeld_2009arXiv0906.1011K}.
This defines the orthogonality condition of the $\eta$-Hermitian eigenvectors of $H$. To introduce $\eta$ in the structure of the inner product, we define
\begin{align}\label{eta_function}
    \ket{E_n^L}&=\eta\ket{E_n^R} ~\text{and} \nonumber \\
    \bra{E_n^L}&=\bra{E_n^R}\eta^{\dagger} ~\text{by complex conjugation. } \nonumber \\
    \bra{E_n^L}&=\bra{E_n^R}\eta, ~\text{~since $\eta$ is Hermitian, see~\eqref{etadefini}. } 
\end{align}
From this, it follows that
\begin{align}
\ket{E_n^L}^{\dagger}&=\bra{E_n^R}\eta^{\dagger}=\bra{E_n^R}\eta=\bra{E_n^L}.
\end{align}
This brings out the $\eta$-metric dependence of the inner product \eqref{InnerP} in a $\eta$-Hermitian vector space
\begin{align}
    \langle{E_m^R}| 
    \eta\ket{E_n^R}=\delta_{mn}.
\end{align}
A state $\ket{E}$ can be written as a weighted linear combination of the complete set of right eigenvectors,
\begin{align}\label{genralstate}
    \ket{E}=\sum_n c_n \ket{E_n^R}.
\end{align}
If $\bra{E_m^L}$ is applied, then by making use of the orthogonality conditions in \eqref{InnerP} we obtain
\begin{align}
    \langle E_m^L \ket{E}&=\sum_n c_n \langle E_m^L\ket{E_n} =\sum_n c_n \delta_{mn} =c_m.
\end{align}
Putting this in \eqref{genralstate}, we obtain (see~\cite{Completeness_orthonormality_PhysRevA.68.062111})
\begin{align}\label{Completeness}
    \ket{E}&=\sum_n \langle  E_n^L\ket{E}~\ket{E_n^R}
    =\left(\sum_n\ket{E_n^R}\bra{E_n^L}\right)~\ket{E}.
\end{align}
This implies the completeness relation
\begin{align}\label{eq_complete}
    &\sum_n\ket{E_n^R}\bra{E_n^L}=\mathbb{I}.
\end{align}
Note that this is equivalent to $\sum_n\ket{E_n^L}\bra{E_n^R}=\mathbb{I}$.
Further, multiplying the completeness relation by the $\eta$ matrix, we get
\begin{align}
    &\sum_n\eta\ket{E_n^R}\bra{E_n^L}=\eta.
\end{align}
From Eq.~\eqref{eta_function}, we have $\eta\ket{E_n^R} = \ket{E_n^L}$, implying
\begin{align}\label{etadefini}
    &\eta=\sum_n \ket{E_n^L}\bra{E_n^L},
\end{align}
from which it follows $\eta^{\dagger}=\sum_n \ket{E_n^L}\bra{E_n^L}=\eta$. Hence $\eta$ is a Hermitian matrix. It can similarly be shown that $\eta^{-1}=(\eta^{-1})^{\dagger}=\sum_n \ket{E_n^R}\bra{E_n^R}$.
Using the completeness relation, we further obtain the spectral decomposition of a general $\eta$-Hermitian Hamiltonian.
\begin{align}\label{SpectralDec}
    H&=IHI \nonumber \\
    &=\sum_{m,n}\ket{E_m^R}\bra{E_m^L}H\ket{E_n^R}\bra{E_n^L} \nonumber \\
    &=\sum_{m,n} E_n\ket{E_m^R}\bra{E_n^L} \delta_{mn} ~~[\text{using \eqref{eigenvalue equation} and \eqref{InnerP}}], \nonumber \\
    &=\sum_{n}E_n \ket{E_n^R}\bra{E_n^L}. 
\end{align}
To explain the $\eta$-Hermiticity of $H$, the adjoint operation in the $\eta$-inner product space is needed.
The complex conjugate of $H$ is $H^{\dagger} =\sum_{n}E_n \ket{E_n^L}\bra{E_n^R}.$ It should be noted that this is not equal to $H =\sum_{n}E_n \ket{E_n^R}\bra{E_n^L}$. The $\eta$-Hermiticity becomes apparent when $\eta^{-1} (\cdot) \eta$ is applied on $H^{\dagger}$ obtaining,
\begin{align}\label{pseudoHermiticity of H}
    \eta^{-1} H^{\dagger} \eta&=\eta^{-1}\left( \sum_{n}E_n \ket{E_n^L}\bra{E_n^R} \right) \eta \nonumber \\
    =&\eta^{-1}\left(\sum_{n}E_n \eta \ket{E_n^R}\bra{E_n^R} \right) \eta = \sum_{n}E_n  \ket{E_n^R}\bra{E_n^R}  \eta \nonumber \\
    =&\sum_{n}E_n  \ket{E_n^R}\bra{E_n^L} = H.
\end{align}
This $\eta$-Hermiticity is the relation connecting $H$ and $H^{\dagger}$ via the $\eta$ metric. This can be used to define the adjoint of any operator $A$ in the $\eta$-inner product Hilbert space 
\begin{align}
    A^{\ddagger}=\eta^{-1} A^{\dagger} \eta,
\end{align}
where $A^{\ddagger}$ is $\eta$-adjoint of the operator $A$.
The $\eta$-Hermiticity of $H$ is then, $H^{\ddagger}=\eta^{-1}H^{\dagger}\eta=H$.

Another way to deal with $\eta$-Hermitian Hamiltonians is by using Hermitian Hamiltonians~\cite{Scolarici2006, Scolarici2007}. A Hermitian matrix $G$ can be found that rotates $H$ to a matrix $H'$, which is Hermitian in nature (see Appendix B)~\cite{kleefeld_2009arXiv0906.1011K},
\begin{align}\label{Htrans}
    H'=GHG^{-1},
\end{align}
where $G^{\dagger}G=G^2=\eta$. 
To calculate $\eta$ and $G$, the right and left eigenvectors of $H$ are needed.
The normalized right eigenvectors of the $\eta$-Hermitian matrix $H$, Eq.~\eqref{nonhermiH},  which is also $\mathcal{PT-}$symmetric are,
\begin{align}\label{righteigs}
    &\ket{E_1^R}=\frac{\ket{e_1^R}}{\sqrt{\langle  e_1^L \ket{e_1^R}}}=\frac{1}{\sqrt{2\cos{\alpha}}}\begin{pmatrix}
        -e^{\frac{-i\alpha}{2}} \\
        e^{\frac{i\alpha}{2}}
    \end{pmatrix}, \nonumber \\
    &\ket{E_2^R}=\frac{\ket{e_2^R}}{\sqrt{\langle  e_2^L \ket{e_2^R}}}=\frac{1}{\sqrt{2\cos{\alpha}}}\begin{pmatrix}
        e^{\frac{i\alpha}{2}} \\
        e^{\frac{-i\alpha}{2}}
    \end{pmatrix}, 
\end{align}
where $\ket{e_n}$'s are the un-normalized eigenvectors of $H$ and $\alpha$ is defined as 
\begin{align}\label{alpha to r s psi}
    \sin{\alpha}=\frac{r}{s}\sin{\psi},
\end{align}
where $r,s$ and $\psi$ are the parameters of $H$. The normalized left eigenvectors of $H$ are equivalent to the normalized right eigenvectors of $H^{\dagger}$ and are
\begin{align}\label{lefteigs}
    &\bra{E_1^L}=\frac{\bra{e_1^L}}{\sqrt{\langle e_1^L \ket{e_1^R}}}=\frac{1}{\sqrt{2\cos{\alpha}}}\begin{pmatrix}
        -e^{\frac{-i\alpha}{2}} & e^{\frac{i\alpha}{2}}\\ 
    \end{pmatrix}, \nonumber \\
    &\bra{E_2^L}=\frac{\bra{e_2^L}}{\sqrt{\langle e_2^L \ket{e_2^R}}}=\frac{1}{\sqrt{2\cos{\alpha}}}\begin{pmatrix}
        e^{\frac{i\alpha}{2}} & e^{\frac{-i\alpha}{2}} \\
    \end{pmatrix} .
\end{align}
Using the above normalized eigenvectors, the metric $\eta$, according to Eq.~\eqref{etadefini}, is 
\begin{align}\label{eta_structure}
    \eta=\begin{pmatrix}
        \sec{\alpha} & -i\tan{\alpha} \\
        i\tan{\alpha} & \sec{\alpha}
    \end{pmatrix}=\eta^{\dagger}.
\end{align}
From this, $G$ is obtained by taking the square root of the matrix $\eta$
\begin{align}\label{Gmatrix}
    G=\frac{1}{\sqrt{\cos{\alpha}}}\begin{pmatrix}
        \cos{\frac{\alpha}{2}} & -i\sin{\frac{\alpha}{2}} \\
        i\sin{\frac{\alpha}{2}} & \cos{\frac{\alpha}{2}}
    \end{pmatrix}.
\end{align}
Using the Hermitian $G$ matrix, $H'=GHG^{-1}$ is obtained as
\begin{align}\label{H'}
    H'=\begin{pmatrix}
        r\cos{\theta} & \sqrt{s^2-r^2\sin^2{\theta}} \\
        \sqrt{s^2-r^2\sin^2{\theta}} & r\cos{\theta}
    \end{pmatrix}.
\end{align}
This Hermitian matrix $H'$ can be written in terms of Pauli matrices ($\sigma_k$ for $k = x,y, z$) as,
\begin{align}
    H'=(r\cos{\theta}) I + (\sqrt{s^2-r^2\sin^2{\theta}}) \sigma_x.
\end{align}
The notion of $\eta$-Hermiticity can be shown to be equivalent to the $\mathcal{CPT}$ formalism. The $\mathcal{CPT}$ inner product is defined for two states $\ket{\xi}$ and $\ket{\zeta}$ as ~\cite{Bender2002}
\begin{align}
    \bra{\zeta}\xi\rangle_{\mathcal{CPT}}=(\mathcal{CPT}\ket{\zeta})^{T}\ket{\xi}.
\end{align}
Here, we have verified that
\begin{align}
    (\mathcal{CPT}\ket{E_m^R})^{T}\ket{E_n^R}=\bra{E_m^L}E_n^R\rangle=\delta_{nm},
\end{align}
when the operator $\mathcal{C}=\mathcal{P}\eta$ \cite{benderbook2019pt}.

\section{Dynamics of an open system for a $\mathcal{PT-}$symmetric Hamiltonian}\label{SecIII}
\begin{figure}
    \centering
    \includegraphics[width=1\linewidth]{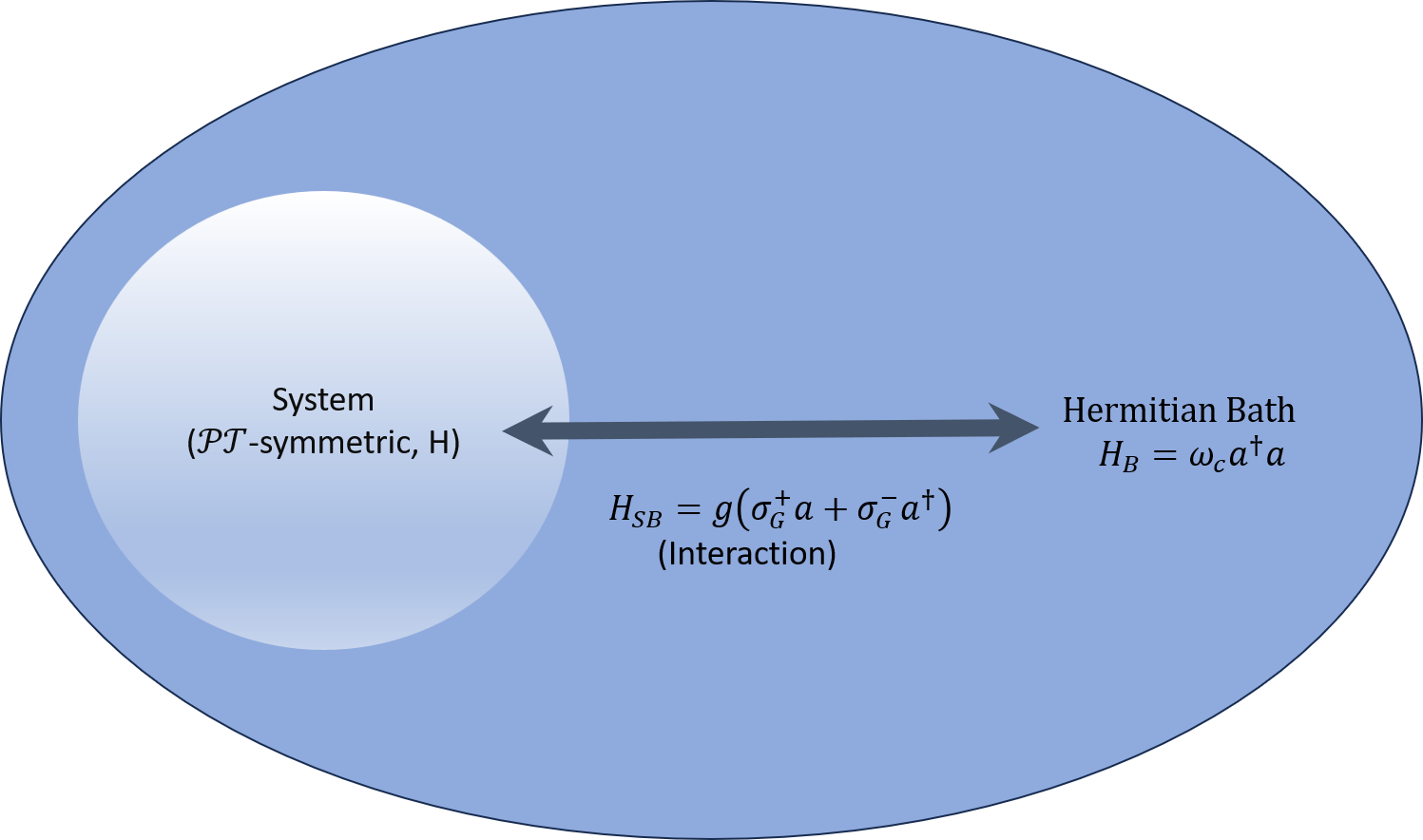}
    \caption{A diagrammatic representation of the open system scheme we apply for a $\pt$symmetric Hamiltonian.}
    \label{fig_PTsymmetry_OpenSystem_diagram}
\end{figure}
We consider a scenario where a single-mode bosonic bath of frequency $\omega_c$ is coupled to the $\mathcal{PT-}$symmetric Hamiltonian $H$, Fig.~\ref{fig_PTsymmetry_OpenSystem_diagram}. Inspired by light-matter interaction models, e.g., the Jaynes-Cummings model, the corresponding composite system-bath Hamiltonian can be envisaged to be
\begin{align}\label{HUtilde}
    \tilde{H}_U&=H + H_B + g(\sigma^+_\mathcal{G}\otimes a + \sigma^-_\mathcal{G} \otimes a^{\dagger}),
\end{align}
where
\begin{align}\label{eq_bath_ham}
    H_B&=\omega_ca^{\dagger}a,
\end{align}
and $g$ is the coupling constant taken to be real.
$\tilde{H}_U$, Eq.~\eqref{HUtilde}, is $\eta$-Hermitian, i.e.,
\begin{align}\label{etahermiofComposite}
    (\eta^{-1}\otimes I_B)\tilde{H}_U^{\dagger}(\eta \otimes I_B)=\tilde{H}_U^\ddagger = \tilde H_U,
\end{align}
following the $\eta$-Hermiticity of $H$, i.e., $\eta^{-1}H^{\dagger}\eta=H$ and the fact that $\sigma^+_\mathcal{G}$ and $\sigma^-_\mathcal{G}$, the system operators suitable for the $\eta$-inner product space,
\begin{align}\label{Biortho_sigmas}
    \sigma^+_\mathcal{G} &=\ket{E_2^R}\bra{E_1^L}, \nonumber \\
    \sigma^-_\mathcal{G} &=\ket{E_1^R}\bra{E_2^L} ,
\end{align}
are $\eta$-adjoints of each other, as shown below
\begin{align}
    (\sigma^+_\mathcal{G})^{\ddagger}&=\eta^{-1}(\sigma^+_\mathcal{G})^{\dagger}\eta =\eta^{-1}\ket{E_1^L}\bra{E_2^R}\eta \nonumber \\
    &=\eta^{-1}\eta\ket{E_1^R}\bra{E_2^L} = \ket{E_1^R}\bra{E_2^L} =\sigma^-_\mathcal{G}.
\end{align}
The corresponding relation for the Hermitian case is $(\sigma^+)^\dagger = \sigma^-$. To understand the dynamics of the $\mathcal{PT-}$symmetric system, we need to obtain the generalized density matrix $\rho_{\mathcal{G}}(t)$ by tracing out the environment from the total $\eta$-unitary evolution $U$ driven by $\tilde{H}_U$, Eq.~\eqref{HUtilde}. It is crucial to apply $U^{\ddagger}$ in the following map
\begin{align}\label{rhoGt_eta_uni}
    \rho_{\mathcal{G}}(t)=\text{Tr}_B[U\left\{\rho_{\mathcal{G}}(0)\otimes \rho_B(0)\right\}U^{\ddagger}],
\end{align}
where the initial system state is taken as the $\mathcal{PT-}$symmetric excited state, $\rho_{\mathcal{G}}(0)=\rho_{\mathcal{G}}^+$, Eq.~\eqref{initialrhoG}, and $\rho_B(0)$ is the initial state of the bath. Here, $\rho_{\mathcal{G}}(t)$ is the density matrix of the $\mathcal{PT-}$symmetric system pertaining to the $\eta$-inner product space.
The $\eta$-adjoint of the $\eta$-unitary operator ~\cite{Scolarici2006,Scolarici2007} is (see Appendix C for the details)
\begin{align}\label{Uddager}
    U^{\ddagger}&=e^{i\tilde{H}_U t}.
\end{align}
The $\rho_{\mathcal{G}}(t)$ computed from Eq.~\eqref{rhoGt_eta_uni} is a legitimate density matrix in $\eta$-inner product space with unit trace. The usual Hermitian density matrix is $\rho(t)=\sum_{ij}\rho_{ij}\ket{E_i}\bra{E_j}$, where $\ket{E_i}\bra{E_j}$s are made by the eigenvectors of the governing Hamiltonian and its complex conjugate. When evolved via a $\mathcal{PT-}$symmetric $H$ (different left and right eigenvectors), $\rho(t)$ does not preserve trace since $H\neq H^{\dagger}$ and the commutator structure of the Schr\"{o}dinger equation is not formed
\begin{align}
    \frac{d}{dt}\rho(t)=-i\left[H\rho(t)-\rho(t) H^{\dagger}\right].
\end{align}
From this, we have
\begin{align}
    \frac{d}{dt}\text{Tr}[\rho(t)]&=-i\text{Tr}\left[H\rho(t)-\rho(t) H^{\dagger}\right] \nonumber \\
    &=-i\text{Tr}\left[(H- H^{\dagger})\rho(t)\right].
\end{align}
In the $\eta$-inner product space, since $H^{\dagger}=\eta H \eta^{-1}$, this implies $\eta H= H^{\dagger}\eta$. $\eta$-Hermiticity of $H$ is leveraged to find bi-orthonormal states with left energy eigenvectors different from the right eigenvectors, Eq.~\eqref{InnerP}. This motivates the definition of generalized density matrix written in terms of the $\eta$-modified projectors
\begin{align}\label{spectral_rhoGt}
    \rho_{\mathcal{G}}(t)&=\sum_{i,j}\rho_{ij}\ket{E_i^R}\bra{E_j^L}\nonumber \\
    &=\sum_{i,j}\rho_{ij}\ket{E_i^R}\bra{E_j^R}\eta=\rho(t)\eta.
\end{align}
Here $\rho(t)=\sum_{i,j}\rho_{ij}\ket{E_i^R}\bra{E_j^R}$ signifies the improper density matrix constructed using the eigenvectors in Eq.~(\ref{righteigs}).
Thus, the above equation can be modified as
\begin{align}
    \frac{d}{dt}\text{Tr}\left[\rho_{\mathcal{G}}(t)\right]&=\text{Tr}\left[\frac{d\rho(t)}{dt}\eta\right]=-i\text{Tr}\left[\left\{H\rho(t)-\rho(t)  H^{\dagger}\right\}\eta\right] \nonumber \\
    &=-i\text{Tr}\left[H\rho(t)\eta-\rho(t)  H^{\dagger}\eta\right] \nonumber \\
    &=-i\text{Tr}\left[H\rho(t)\eta-\rho(t) \eta H\right] \nonumber \\
    &=-i\text{Tr}\left[H\rho_{\mathcal{G}}(t)-\rho_{\mathcal{G}}(t) H\right]=0.
\end{align}
This also shows us that the generalized density matrix obeys Schr\"{o}dinger's equation with the $\mathcal{PT-}$symmetric $H$, that is, 
\begin{align}\label{eq_gen_von_neumann_eq}
    \frac{d}{dt}\rho_{\mathcal{G}}(t)&=-i[H,\rho_{\mathcal{G}}(t)].
\end{align}

The $\rho_{ij}$'s in Eq.~(\ref{spectral_rhoGt}) signifies the population and coherence term between the energy levels $\ket{E_1}$ and $\ket{E_2}$. 
The trace of this generalized density matrix, i.e., action of $\bra{E_k^L}(\cdot)\ket{E_k^R}$, gives $\sum_k\rho_{kk}=1$. In contrast, $\rho(t)=\sum_{ij}\rho_{ij}\ket{E_i}\bra{E_j}$ remains an improper density matrix because $\bra{E_i}E_j\rangle\neq \delta_{ij}$ and the trace of this produces a non-conserved quantity that involves the parameters in the Hamiltonian $H$ and the coherence terms of $\rho(t)$.

Another approach to finding $\rho_{\mathcal{G}}(t)$ is to find a Hermitian Hamiltonian of the form, Eq.~\eqref{H'}. 
Using the property of the Hermitian matrix $G$, $\eta=G^2$, we arrive at Eq.~\eqref{Htrans}, discussed in Appendix B.
The spectral decomposition of $H'$ is,
\begin{align}\label{eq_H_prime}
    H'&=GHG^{-1} \nonumber \\
    &=G(\sum_n E_n \ket{E_n^R}\bra{E_n^L})G^{-1} \nonumber \\
    &=\sum_n E_n G\ket{E_n^R}\bra{E_n^R} \eta G^{-1} \nonumber \\
    &=\sum_n E_n G\ket{E_n^R}\bra{E_n^R} G .
\end{align}
This shows how the projectors are modified by the above operation with the completeness relation $I=\sum_nG\ket{E_n^R}\bra{E_n^R}G$, which follows from Eq.~\eqref{eq_complete} and $G^2 = \eta$. 
Evidently, this Hermitian space is endowed with the same right and left eigenvectors, $G\ket{E_n^R}$ and $\bra{E_n^R}G$, obeying the orthogonality condition of the $\eta$-inner product space,
\begin{align}
    \bra{E_i^R}G~G\ket{E_j^R}=\bra{E_i^R}\eta\ket{E_j^R}=\bra{E_i^L}E_j^R\rangle=\delta_{ij}.
\end{align}
The Hermitian density matrix $\rho'(t)$, obtained by the evolution of $H'$, Eq.~\eqref{eq_H_prime}, is defined by the modified projectors as
\begin{align}
    \rho'(t)=\sum_{ij}\rho_{ij}G\ket{E_i^R}\bra{E_j^R}G.
\end{align}
From the above decomposition, the relation of $\rho'(t)$ with $\rho_{\mathcal{G}}(t)$ and $\rho(t)$ is evident
\begin{align}
    G^{-1}\rho'(t)G=\sum_{ij}\rho_{ij}\ket{E_i^R}\bra{E_j^R}\eta=\rho_{\mathcal{G}}(t), \nonumber \\
    G^{-1}\rho'(t)G^{-1}=\sum_{ij}\rho_{ij}\ket{E_i^R}\bra{E_j^R}=\rho(t).
\end{align}
The corresponding system operators in the Hermitian space are $G\ket{E_2^R}\bra{E_1^R}G=\sigma'^{+}$ and $G\ket{E_1^R}\bra{E_2^R}G=\sigma'^{-}$. They are connected to the system operators in the $\eta$-inner product space as
\begin{align} \label{sigmaGplus}
    \sigma_{\mathcal{G}}^+=G^{-1}(G\ket{E_2^R}\bra{E_1^R}G)G=G^{-1}(\sigma'^{+})G, \nonumber \\
    \sigma_{\mathcal{G}}^-=G^{-1}(G\ket{E_1^R}\bra{E_2^R}G)G=G^{-1}(\sigma'^{-})G.
\end{align}

Now, we come back to the open system discussion from the perspective of this Hermitian space. Since $\tilde{H}_U$ in Eq.~\eqref{HUtilde} is non-Hermitian because of the presence of $H$ in it, it cannot drive a unitary evolution. To this end, a Hermitian Hamiltonian $H_U$ can be obtained by applying $(G\otimes I_B)(\tilde H_U)(G\otimes I_B)^{-1}$, that yields
\begin{align}\label{HUforPT}
    H_U&=GHG^{-1}\otimes I_B + G\cdot G^{-1}\otimes H_B \nonumber \\
    &+ g(G \sigma_{\mathcal{G}}^+ G^{-1} \otimes a + G\sigma_{\mathcal{G}}^- G^{-1} \otimes a^{\dagger}) \nonumber \\
    &=H'\otimes I_B + I_2\otimes H_B \nonumber \\
    &+ g( \sigma'^{+} \otimes a + \sigma'^{-}  \otimes a^{\dagger}), \nonumber \\
    &=H'+H_B+g(\sigma'^{+}\otimes a+\sigma'^{+} \otimes a^{\dagger}).
\end{align}
where Eq.~\eqref{sigmaGplus} has been used.
We now have a composite Hermitian Hamiltonian $H_U$ suitable for unitary evolution $U' = e^{-iH_Ut}$ using which $\rho'(t)$ is obtained by
\begin{align}
    \rho'(t)=\text{Tr}_B[U'\left\{\left(G\rho_{\mathcal{G}}(0)G^{-1}\right)\otimes \rho_B(0)\right\}U'^{\dagger}].
\end{align}
Then we transform it back to the space of the $\eta$-Hermitian Hamiltonian $H$ using the $G$ matrix again, i.e.,
\begin{align}\label{rotation_of_rho}
    \rho_{\mathcal{G}}(t)=G^{-1}\rho'(t)G.
\end{align}

In the above Eq.~\eqref{rotation_of_rho} and in Eq.~\eqref{rhoGt_eta_uni} we outlined two prescriptions to obtain $\rho_{\mathcal{G}}(t)$. Both prescriptions provide the same results for the quantities studied in the subsequent sections. The method illustrated by Eq.~\eqref{rotation_of_rho} involving $\rho'(t)$ and Hermitian $H'$ is suited for dealing with a master equation endowed with a Hermitian Hamiltonian and Lindblad jump operators~\cite{Scolarici2006}.

To proceed with the analysis of the dynamics of $\rho_{\mathcal{G}}(t)$, the ground and excited states of $\rho_{\mathcal{G}}(t)$ under the non-Hermitian evolution under $H$ must be ascertained.
We have the right and left eigenvectors of $H$ in \eqref{righteigs} and \eqref{lefteigs}. It is evident from the real eigenvalues $E_1$ and $E_2$ that $E_1 < E_2$. Thus, we define $\ket{E_1^R}\bra{E_1^L}$ as the ground state and $\ket{E_2^R}\bra{E_2^L}$ as the excited state in terms of generalized density matrices. In terms of the parameters of the Hamiltonian $H$, we obtain these states as
\begin{align}\label{initialrhoG_1}
    \rho_{\mathcal{G}}^-=\ket{E_1^R}\bra{E_1^L}&=\frac{1}{2}\begin{pmatrix}
    1 - i \tan \alpha & -\sec \alpha \\
    -\sec \alpha & 1 + i \tan \alpha
    \end{pmatrix},
\end{align}
\begin{align}\label{initialrhoG}
    \rho_{\mathcal{G}}^+=\ket{E_2^R}\bra{E_2^L}&=\frac{1}{2} \begin{pmatrix}
    1 + i \tan \alpha & \sec \alpha \\
    \sec \alpha & 1 - i \tan \alpha
    \end{pmatrix} .
\end{align}
The value of $\alpha$ is determined by Eq.~\eqref{alpha to r s psi}. The values $r=0.1$, $s=0.4$, and $\psi=\pi/6$ are taken throughout the paper.

Here, we have constructed a generalized density matrix $\rho_{\mathcal{G}}(t)$ that is suitable for a quantum evolution driven by a $\pt$symmetric Hamiltonian. The structure of the generalized density matrix proposed, i.e., $\rho_{\mathcal{G}}(t)=\sum_{i,j}\rho_{ij}\ket{E_i^R}\bra{E_j^L}$ differs from a usual density matrix for an evolution by a Hermitian Hamiltonian. The bi-orthonormal eigenspace accommodates the information elements of a quantum state, i.e., the population and coherence terms ($\rho_{ij}$'s). In this non-Hermitian setup, the appropriate generalized density matrix contains the parameters of the $\pt$symmetric Hamiltonian in addition to the population and coherence terms typical of a quantum state. We are interested in analyzing some information theoretic measures using the generalized density matrix $\rho_{\mathcal{G}}(t)$ in order to observe how the $\pt$symmetric geometry affects the information content of the system.

The GKSL master equation provides a very well-known description of open system evolutions, albeit in the weak coupling, Markovian approximation. Our goal in this paper is to examine a $\pt$symmetric open quantum system without any approximation. However, keeping in mind the wide usage of the GKSL master equation, we provide a corresponding derivation in Appendix~\ref{sec_der_pt_gksl} for the $\pt$symmetric system.

\section{Information-theoretic study of a non-Hermitian open system: comparison with its Hermitian counterpart}\label{SecIV}
We take a two-level Hermitian Hamiltonian of a single spin $H_0$, and couple it to a bath Hamiltonian $H_B=\omega_ca^{\dagger}a$ with a coupling similar to Eq.~\eqref{HUtilde}, albeit with the usual atomic raising and lowering operators $\sigma_+ = (\sigma_x + i\sigma_y)/2$ and $\sigma_- = (\sigma_x - i\sigma_y)/2$, respectively. The corresponding system-bath Hamiltonian is given by
\begin{align}\label{H0U}
    H_U^0&=H_0\otimes I_B + I_2\otimes H_B + g( \sigma_+ \otimes a + \sigma_-  \otimes a^{\dagger}),
\end{align}
where 
\begin{align}
    H_0=\omega_0\sigma_z.
\end{align}
We obtain the system density matrix $\rho_S(t)$ numerically by tracing out the bath degrees of freedom from the total unitary evolution by $H_U^0$ as
\begin{align}
    \rho_S(t)=\text{Tr}_B[e^{-iH^0_U t}(\rho_S(0)\otimes \rho_B)e^{iH^0_U t}].
\end{align}
We calculate a number of information-theoretic quantities using $\rho_{\mathcal{G}}(t)$ for a non-Hermitian $\mathcal{PT-}$symmetric system Hamiltonian and $\rho_S(t)$ for a Hermitian system Hamiltonian and make a comparison.

Here, we use the method of directly partial tracing out the Hermitian bath degrees of freedom. This generalized method takes no approximation into account and can include any arbitrary value of the coupling constant $g$. This allows us to work with strong coupling scenarios beyond the Markovian regime. This gives rise to non-Markovian effects, which could be witnessed here by the revival characteristics, identified by the BLP measure.

The non-Markovian nature of the above evolution is identified by using the BLP measure. Further, we compare the fidelity of the states evolved using the non-Hermitian $\mathcal{PT-}$symmetric system Hamiltonian with the fidelity of the states evolved using a Hermitian system.  
\subsection{The BLP measure}
\begin{figure}
    \centering
    \includegraphics[width=1\linewidth]{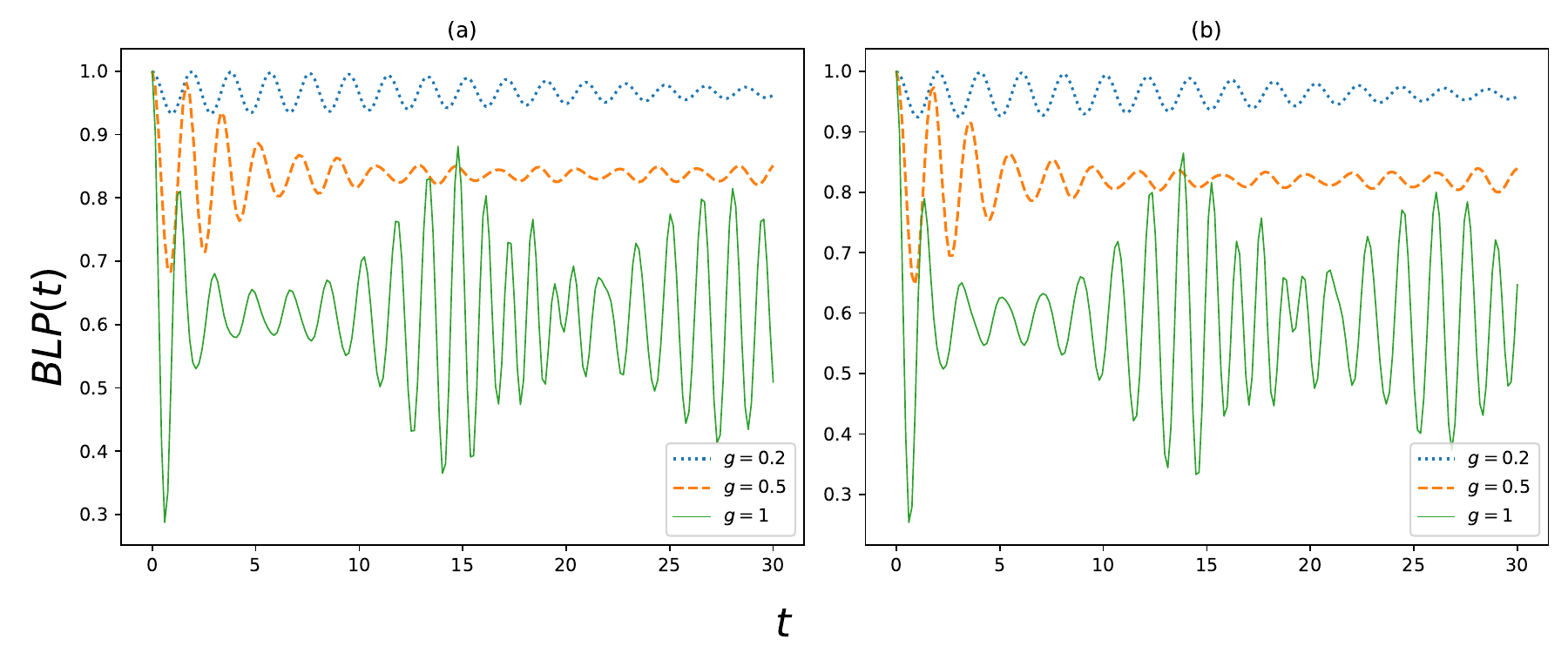}
    \caption{The trace distance is plotted in (a) for the dynamics of $\mathcal{PT-}$symmetric system Hamiltonian $H$ and (b) for the dynamics of Hermitian $H_0$. The two initial states in (a) are the ground state $\rho_{\mathcal{G}}^-$ and the excited state $\rho_{\mathcal{G}}^+$, whereas in (b) the initial states are the ground and the excited states of a two-level Hermitian spin system. The other parameters are $T=10$, $\omega_c=2$, dimension of bath = 10.}
    \label{fig:BLP_compare_Pt_nonPT}
\end{figure}

The Breuer-Laine-Piilo (BLP) measure is related to the time evolution of the trace distance between two different states as a measure of non-Markovianity~\cite{lainebreuer}. For generalized density matrices, the trace distance
\begin{align}
    &D(\rho_{\mathcal{G}1}(t), \rho_{\mathcal{G}2}(t)) \nonumber \\
    &= \frac{1}{2}{\rm Tr}\sqrt{(\rho_{\mathcal{G}1}(t) - \rho_{\mathcal{G}2}(t))^\ddagger(\rho_{\mathcal{G}1}(t) - \rho_{\mathcal{G}2}(t))},
\end{align}
is calculated between two states $\rho_{\mathcal{G}1}(t)$ and $\rho_{\mathcal{G}2}(t)$ of the system at any time $t$. Note that the implementation of the $\eta$-adjoint is crucial when dealing with an $\eta$-inner product space. Revivals in the variation of this trace distance are considered to be the marker of non-Markovianity of the evolution. For the BLP measure of $\rho_S(t)$, the adjoint is simply the complex conjugate.
We compute the trace distance for both $\rho_{\mathcal{G}}(t)$ and $\rho_S(t)$ for different values of the coupling constant $g$. The initial states of the $\mathcal{PT-}$symmetric generalized density matrix are $\rho_{\mathcal{G}}(0)^1=\rho_{\mathcal{G}}^+$ and $\rho_{\mathcal{G}}(0)^2=\rho_{\mathcal{G}}^-$, given in Eqs.~\eqref{initialrhoG_1} and~\eqref{initialrhoG}. The initial states for the Hermitian spin system are
\begin{align}
    \rho_S(0)^1&=\rho_S^+ = \begin{pmatrix}
    1 & 0 \\
    0 & 0
    \end{pmatrix},\\
    \rho_S(0)^2&=\rho_S^- = \begin{pmatrix}
    0 & 0 \\
    0 & 1
    \end{pmatrix}.
\end{align}
Both sets of chosen initial states are the excited and ground states.
The initial state of the bath is taken to be
$
    \rho_B(0)=e^{-H_B/T}/{\rm Tr}(e^{-H_B/T}),
$ where $T$ is the temperature. 
The comparisons of the trace distance are depicted in Fig.~\ref{fig:BLP_compare_Pt_nonPT}.
We observe that the trace distance for both the cases of the non-Hermitian $\mathcal{PT-}$symmetric system and that of the Hermitian system shows revivals in trace distance, indicating the non-Markovian evolution of the system. We have taken the frequency $\omega_0=0.5$ for the Hermitian $H_0$ and $r=0.1,~s=0.4,~\psi=\pi/6$ for the $\pt$symmetric $H$ to make the matrix elements comparable in magnitude. As a result, we obtain a qualitatively similar profile of BLP as seen in the figure, despite the fact that the dynamics are governed by Hamiltonians that are of two entirely different types, i.e.,  $\pt$symmetric and Hermitian. Next, we check the fidelity.
\subsection{Fidelity}
In quantum information theory, fidelity is a measure of the ``closeness" between two density matrices. It ranges from 
0 (completely distinguishable states) to 1 (identical states).
Here, we analyze the fidelity between the final and the initial $\mathcal{PT-}$symmetric states and compare it with that of a two-level Hermitian spin system. The fidelity is given by
\begin{equation}
F(\rho_{\mathcal{G}}(t), \rho_{\mathcal{G}}(0)) = \left( \text{Tr} \sqrt{\sqrt{\rho_{\mathcal{G}}(t)} \rho_{\mathcal{G}}(0) \sqrt{\rho_{\mathcal{G}}(t)}} \right)^2.
\end{equation}
Here, $\sqrt{\rho_{\mathcal{G}}(t)} \rho_{\mathcal{G}}(0) \sqrt{\rho_{\mathcal{G}}(t)}$ is a positive semidefinite density matrix.
We compare the fidelity growth $F(t)$ of the $\rho_{\mathcal{G}}(t)$ and $\rho_S(t)$ with their corresponding initial states to gain insight into the $\mathcal{PT-}$symmetric system. The initial states taken are $\rho_{\mathcal{G}}(0)=\rho_{\mathcal{G}}^+$ and $\rho_S(0)=\rho_S^+$.

\begin{figure}
    \centering
    \includegraphics[width=1\linewidth]{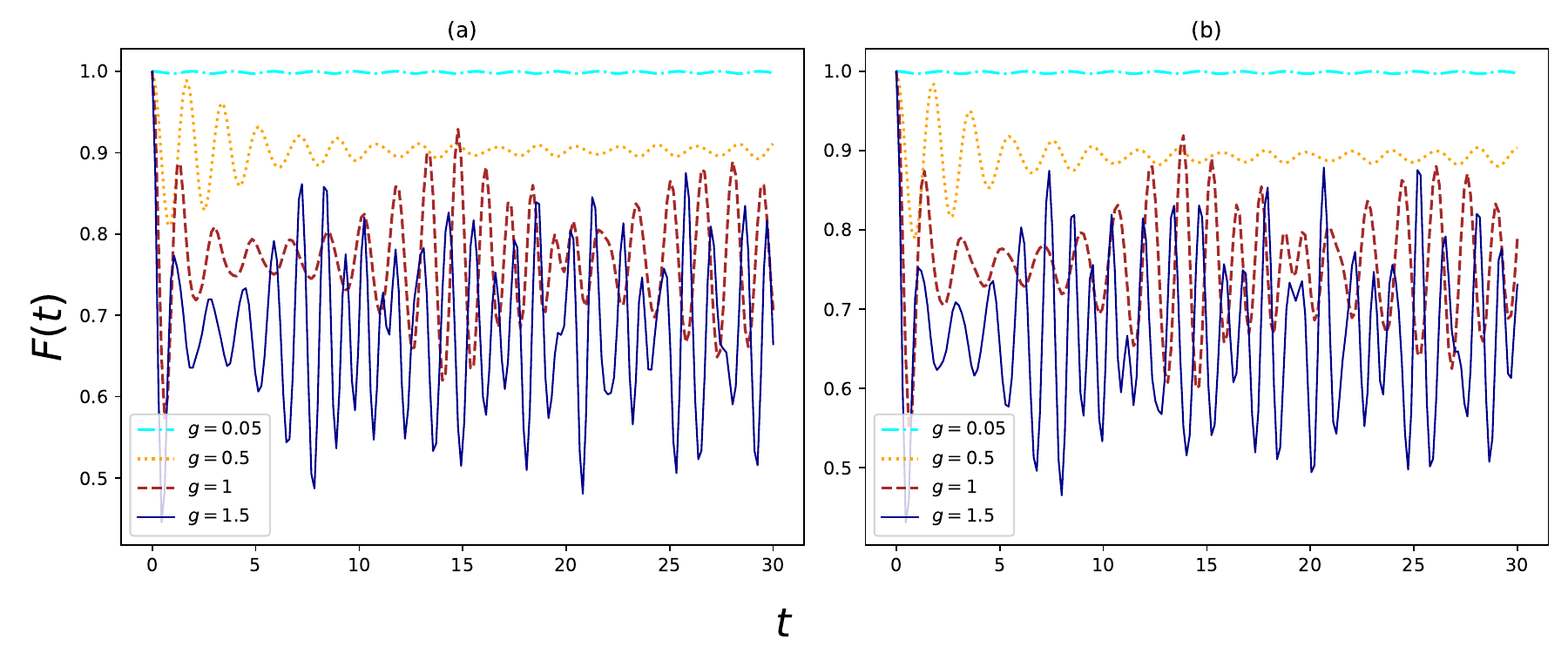}
    \caption{Variation of the fidelity $F(t)$ with different values of coupling constants $g$ is shown for the dynamics of $\mathcal{PT-}$symmetric $H$ in (a) and for the dynamics of Hermitian $H_0$ in (b). Other parameters are $T=10$, $\omega_c=\omega_0 = 2$.  }
    \label{fig:Fidelity_gcompare_Pt_nonPT}
\end{figure}
\begin{figure}
    \centering
    \includegraphics[width=1\linewidth]{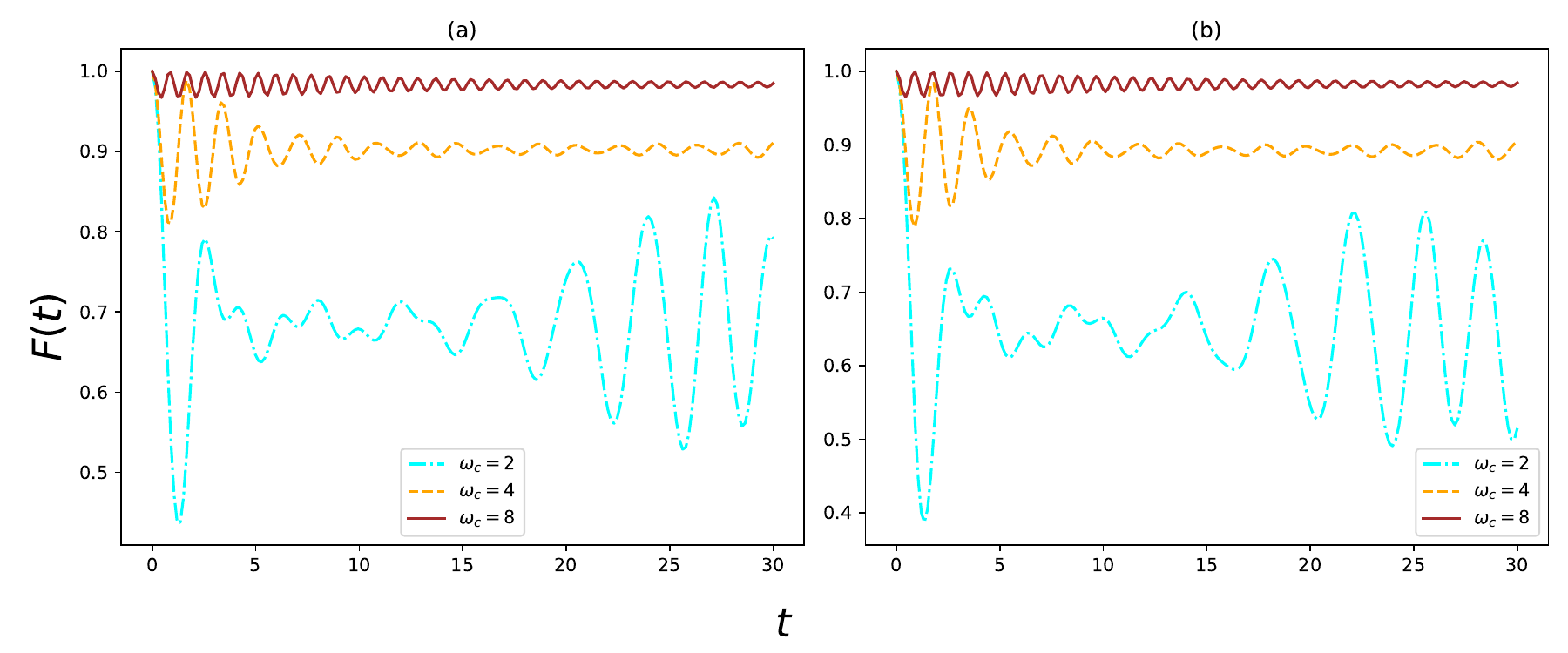}
    \caption{Variation of the fidelity $F(t)$ with different values of the bath frequency $\omega_c$ is shown for the dynamics of $\mathcal{PT-}$symmetric $H$ in (a) and for the dynamics of Hermitian $H_0$ in (b). Other parameters are $T=10$, $g=0.5$, $\omega_0 = 2$.}
    \label{fig:Fidelity_wccompare_Pt_nonPT}
\end{figure}
\begin{figure}
    \centering
    \includegraphics[width=1\linewidth]{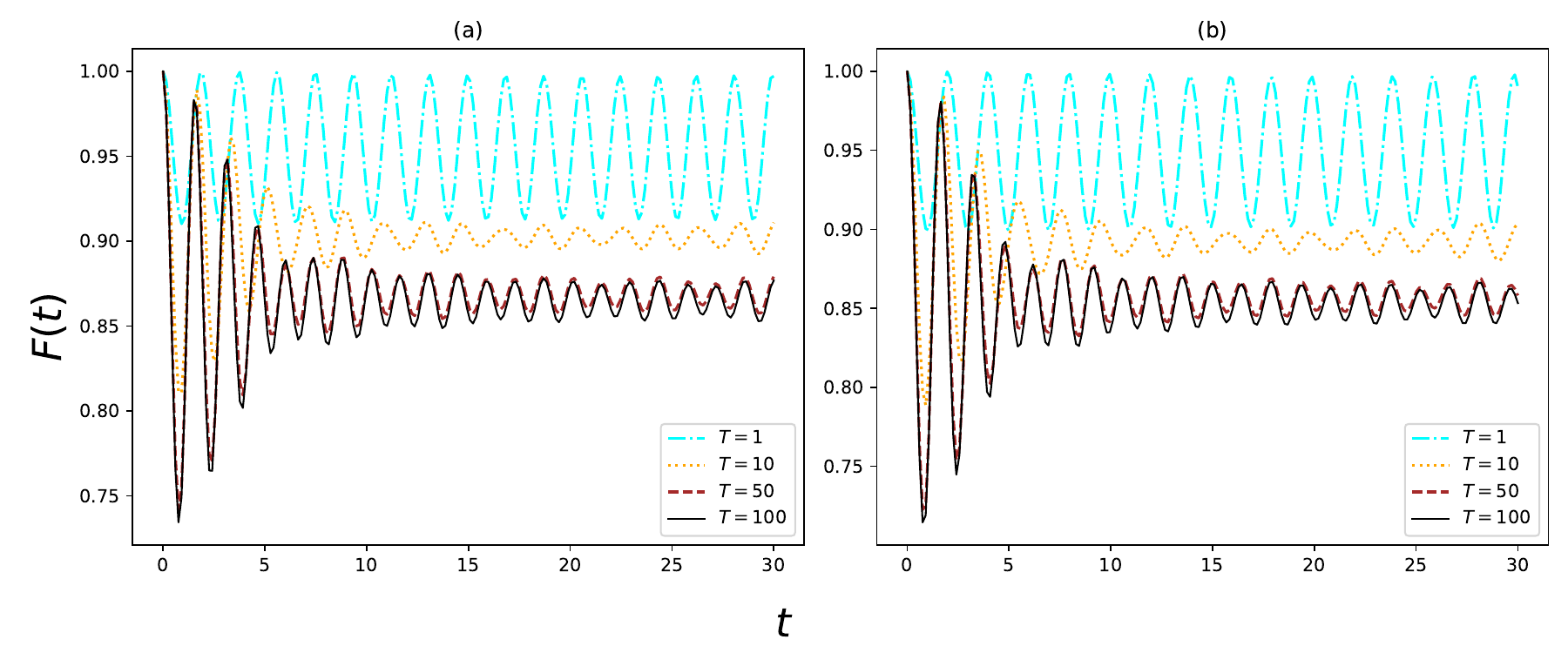}
    \caption{Variation of the fidelity $F(t)$ with different values of temperature of the bath $T$ is shown for the dynamics of $\mathcal{PT-}$symmetric $H$ in (a) and for the dynamics of Hermitian $H_0$ in (b). Other parameters are $g=0.5$, $\omega_c=\omega_0 = 2$.}
    \label{fig:Fidelity_Tcompare_Pt_nonPT}
\end{figure}

In Figs.~\ref{fig:Fidelity_gcompare_Pt_nonPT},~\ref{fig:Fidelity_wccompare_Pt_nonPT} and~\ref{fig:Fidelity_Tcompare_Pt_nonPT}, comparisons of fidelity for different coupling constants g, bath frequencies $\omega_c$ and bath temperature $T$ are depicted for both kinds of Hamiltonians. Again, it is observed that the fidelity $F(t)$ for the $\pt$symmmetric and the Hermitian evolution starting from their corresponding excited states behave qualitatively in a similar way across all parameter ranges. 
We now observe whether the entanglement generation also preserves this similarity for these two distinct quantum evolutions.

\section{Entanglement between two $\mathcal{PT-}$symmetric open systems}\label{SecV}
To ascertain the development of entanglement between two $\mathcal{PT-}$symmetric systems over time, we take two systems having the same $\mathcal{PT-}$symmetric Hamiltonian $H$ with a nearest neighbor interaction $\sigma_{\mathcal{G}}^z-\sigma_{\mathcal{G}}^z$, and interacting with the bath $H_B$. The parameters of $H$ are kept the same as before, i.e., $r=0.1$, $s=0.4$, and $\psi=\pi/6$. The Hamiltonian of the total evolution ($\eta$-unitary) is
\begin{align}
    \tilde{H}_U^e&=H^{(1)}\otimes I_2+I_2\otimes H^{(2)} +j(\sigma_{\mathcal{G}}^{z,(1)}\otimes \sigma_{\mathcal{G}}^{z,(2)})\nonumber \\
    &+H_B+g\{ (\sigma^{+,(1)}_{\mathcal{G}}\otimes I_2 + I_2\otimes \sigma^{+,(2)}_{\mathcal{G}})\otimes a \nonumber \\
    &+ (\sigma^{-,(1)}_{\mathcal{G}} \otimes I_2 + I_2 \otimes \sigma^{-,(2)}_{\mathcal{G}} )\otimes a^{\dagger}\}, 
\end{align}
where Hamiltonians $H^{(1)}$ and $H^{(2)}$ have the form of Eq.~\eqref{nonhermiH} and $H_B$ is the same as Eq.~\eqref{eq_bath_ham}.
$\rho_{\mathcal{G}}(t)$ of the composite system can be obtained by Eq.~\eqref{rhoGt_eta_uni} with $U^{\ddagger}=e^{i\tilde{H}^e_U t}$ where the system's initial state is $\rho_{\mathcal{G}}^+\otimes \rho_{\mathcal{G}}^+$.

\begin{figure}
    \centering
    \includegraphics[width=1\linewidth]{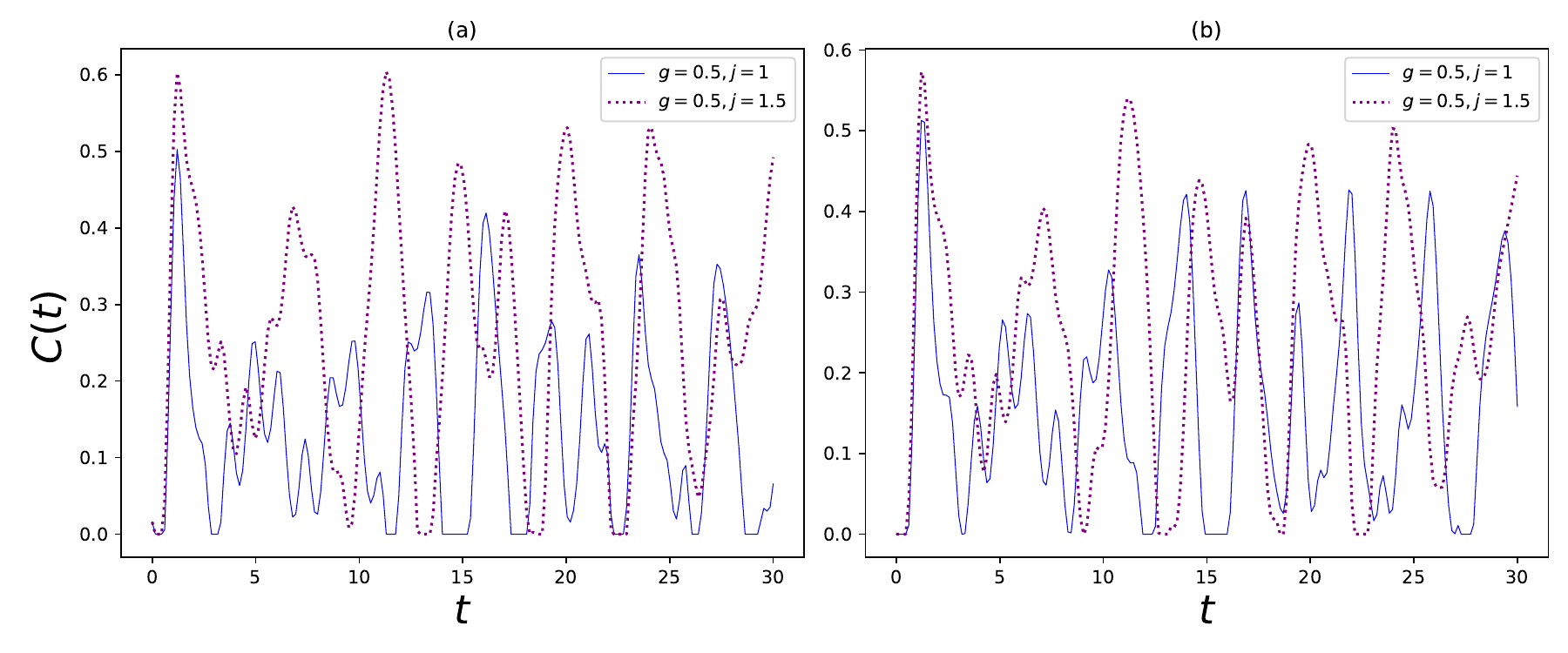}
    \caption{Concurrence of the two systems coupled to each other with coupling strength $j$ subjected to an environment with coupling g is shown for the dynamics of $\mathcal{PT-}$symmetric $H$ in (a) and for the dynamics of Hermitian $H_0$ in (b). The other parameters are $T=10$, $\omega_c=4$.}
    \label{fig:Concurrence_Pt}
\end{figure}
To quantify entanglement, we calculate concurrence~\cite{Wootters1998, Pati2009, Scolarici_2009}.
Given a composite generalized density matrix for two $\mathcal{PT-}$symmetric systems \(\rho_{\mathcal{G}}(t)\), the concurrence \(C\left[\rho_{\mathcal{G}}(t)\right]\) is defined as
\begin{equation}
    C\left[\rho_{\mathcal{G}}(t)\right] = \max \left( 0, \lambda_1 - \lambda_2 - \lambda_3 - \lambda_4 \right),
\end{equation}
where \(\lambda_i\) are the square roots of the eigenvalues, in decreasing order, of the non-Hermitian matrix
\begin{equation}
    R = \rho_{\mathcal{G}}(t) \, \tilde{\rho}_{\mathcal{G}}(t),
\end{equation}
with
\begin{equation}
    \tilde{\rho}_{\mathcal{G}}(t) = \left( \sigma_{\mathcal{G}}^y \otimes \sigma_{\mathcal{G}}^y \right) \rho_{\mathcal{G}}^*(t) \left( \sigma_{\mathcal{G}}^y \otimes \sigma_{\mathcal{G}}^y \right).
\end{equation}
Here, \(\rho^*_{\mathcal{G}}(t)\) denotes the complex conjugate of \(\rho_{\mathcal{G}}(t)\) , and \(\sigma_{\mathcal{G}}^y\) is the Pauli-$Y$ matrix on the bi-orthonormal basis, redefined as, 
\begin{align}\label{biorthoSigmas}
    \sigma^x_\mathcal{G} &=\ket{E_1^R}\bra{E_2^L}+\ket{E_2^R}\bra{E_1^L}, \nonumber \\
    \sigma^y_\mathcal{G} &=i\ket{E_1^R}\bra{E_2^L}-i\ket{E_2^R}\bra{E_1^L} , \nonumber \\
    \sigma^z_\mathcal{G} &=\ket{E_2^R}\bra{E_2^L}-\ket{E_1^R}\bra{E_1^L}.
\end{align}
These redefined bi-orthonormal $\sigma_{\mathcal{G}}$ operators obey the same commutation relations as the usual $\sigma$ operators in the computational basis, i.e., $[\sigma_{\mathcal{G}}^i, \sigma_{\mathcal{G}}^j] = 2i \, \varepsilon^{ijk} \, \sigma_{\mathcal{G}}^k$ (See Appendix E).
 We compute \(C\left[\rho_{\mathcal{G}}(t)\right]\) for the evolved states in our \(\mathcal{PT}\)-symmetric setup to capture the dynamical behavior of bipartite entanglement under non-Hermitian dynamics. The concurrence serves as a clear and compact diagnostic tool to identify regimes of high entanglement and to assess the impact of the \(\mathcal{PT-}\)symmetry on quantum correlations.

To compare this result with that of a Hermitian evolution, we calculate $C(t)$ for two qubits each evolving under a Hermitian Hamiltonian $H_0=\omega_0\sigma_z$ and with a total unitary evolution governed by the Hamiltonian
\begin{align}\label{HU0}
    H_U^{0,2}&= H_0^{(1)}\otimes I_2+I_2\otimes H_0^{(2)} +j(\sigma^{z, (1)}\otimes \sigma^{z, (2)}) \nonumber \\
    &+H_B+g\{ (\sigma^{+, (1)}\otimes I_2 + I_2\otimes \sigma^{+, (2)})\otimes a \nonumber \\
    &+ (\sigma^{-, (1)} \otimes I_2 + I_2 \otimes \sigma^{-, (2)} )\otimes a^{\dagger}\}.
\end{align}

Figure~\ref{fig:Concurrence_Pt} shows qualitatively similar irregular oscillations of concurrence $C(t)$ at $j = 1$ and $j=1.5$ for both the $\pt$symmetric and the Hermitian Hamiltonians. The bath parameters that elicit the most clear profile of $C(t)$ are $T=10$, $\omega_c=4$, and $g=0.5$.

The similarity between the concurrence of two entirely different classes of Hamiltonians, i.e., the $\pt$symmetric and the Hermitian spin, indicates that even the entanglement generation does not depend on the geometry of the eigenbasis that constructs the corresponding density matrices. More generally, given the same parameters of the bath and the specific nature of interaction, Hamiltonians of a Hermitian spin system and a $\pt$symmetric system with comparable values of matrix elements result in similar BLP measure, fidelity, and entanglement generation (concurrence) profiles when starting from an excited state (for $H_0$) and its bi-orthonormal version (for $H$). In the bi-orthonormal basis, the quantum state elements behave the same way as the Hermitian systems do in their orthonormal computational basis under a common CPTP evolution. The trace operation necessary for all the information theoretic content releases the quantum elements from the folds of the $\pt$symmetric geometry of the eigenbasis. That is why the information-theoretic quantities bring out the core elements (population, coherence) of the quantum dynamics driven by two different classes of Hamiltonians of comparable magnitude that appear the same information-wise. 
The motivation to perform an information theoretic analysis of a $\pt$symmetric system was provided by the unique structure of the generalized density matrix ($\rho_{\mathcal{G}}(t)=\sum_{i,j}\rho_{ij}\ket{E_i^R}\bra{E_j^L}$). We have seen that not only the generalized density matrix which encodes the quantum state of a $\pt$symmetric Hamiltonian, but also all the system operators in the bi-orthonormal eigenspace of the $\pt$symmetric Hamiltonian contains parameters of the Hamiltonian by means of modified bi-orthonormal projectors which in other words is the artifact of the specific geometry of the system characterized by the $\eta$ metric. Our aim is to develop an open quantum system framework for $\mathcal{PT}-$symmetric quantum systems. To do this, we start with a simple open system model as an example and analyze its dynamics. We also compare the information-theoretic dynamics of a $\pt$symmetric system and a Hermitian spin-half system. What we find is that, despite the Hamiltonians being substantially different, any information-theoretic difference is masked by the CPTP map acting on it. This manifests in the qualitative similarity of the plots for measures like BLP, fidelity, and concurrence.
This work is an open system extension to $\pt$symmetric quantum systems. The key point is to couple a $\pt$symmetric system Hilbert space with a Hermitian bath Hilbert space and then evolving the composite Hilbert space by an $\eta$-unitary evolution to obtain the reduced dynamics of the system. This systematic approach is developed here, and we would extend this work using more complex models in the future.

\section{Conclusions}\label{SecVI}
In this paper, we studied a $\mathcal{PT-}$symmetric Hamiltonian from the perspective of an open quantum system. The $\eta-$Hermiticity of a $\mathcal{PT-}$symmetric Hamiltonian was examined in detail. A first-principles derivation of the generalized GKSL master equation for the $\pt$symmetric system was provided, revealing that its structure closely resembles that of the GKSL master equation for a dissipative two-level system. This was followed by the open system dynamics of this Hamiltonian along the lines of the Jaynes-Cummings type interaction using system operators suitable for $\eta$-inner product space. Two prescriptions were provided to obtain the generalized reduced density matrix of the system. In the first, the field modes were traced out numerically after applying the $\eta$-adjoint unitaries. The other description used a Hermitian Hamiltonian obtained from the original $\eta$-Hermitian Hamiltonian by a transformation using a matrix constructed from the $\eta$ matrix. Both prescriptions were consistent with each other. To benchmark the generalized density matrix obtained from the $\mathcal{PT}$-symmetric Hamiltonian with its Hermitian counterpart, their information-theoretic content was studied. This involved a non-Markovianity witness, that is, the BLP measure, and quantum fidelity. A comparison clearly brought out the qualitative similarity between the $\mathcal{PT-}$symmetric evolution and its Hermitian counterpart. We further studied two interacting $\mathcal{PT-}$symmetric systems, coupled to a bath. Using the procedure outlined above, the generalized reduced state of the system was obtained. This was then used to calculate concurrence between the two $\mathcal{PT-}$symmetric systems. Interestingly, it was found that even the entanglement generation of the $\pt$symmetric system matches that of its Hermitian counterpart. This highlights the fact that the information content (population, coherence) of the corresponding density matrices does not get affected by the bi-orthonormal geometry ($\eta$-Hermiticity) of the $\pt$symmetric basis. In other words, the quantum information evolves according to the nature of the CPTP map applied to the quantum state.

\section*{Acknowledgments}
SB acknowledges Tanmoy Das for useful discussions on $\mathcal{PT}-$symmetry. 

\appendix

\section{The left eigenvectors}
From Eq.~\eqref{eigenvalue equation}, we have,
\begin{align}\label{lefteigenvecs}
    &H\ket{E_n^R}=E_n\ket{E_n^R}.
\end{align}
Taking the complex conjugate and using the $\eta$-Hermiticity
\begin{align}
    \bra{E_n^R}H^{\dagger}=\bra{E_n^R}\eta H\eta^{-1}=E_n\bra{E_n^R}
\end{align}
is obtained and using $\bra{E_n^R}\eta=\bra{E_n^L}$, we have
\begin{align}
    \bra{E_n^L} H=E_n\bra{E_n^L}.
\end{align}
By complex conjugation, we finally obtain
\begin{align}
    H^{\dagger}\ket{E_n^L}=E_n\ket{E_n^L}.
\end{align}
This tells us that the left eigenvectors are the eigenvectors of $H^{\dagger}$.
\section{The Hermitian Hamiltonian}
Starting from the $\eta$-Hermiticity of $H$, $H^{\ddagger}=\eta^{-1}H^{\dagger}\eta =H$, we can decompose $\eta$ as $\eta=G^2$ as
\begin{align}\label{bracketHdag}
    H=G^{-1}(G^{-1}H^{\dagger}G)G.
\end{align}
If  $G^{-1}H^{\dagger}G=H'$, then we have,
\begin{align}
    &H=G^{-1}H'G.
\end{align}
The definition of $H'$ below Eq.~\eqref{bracketHdag} assures the Hermiticity of $H'$. Taking complex conjugate of $H'=G^{-1}H^{\dagger}G$, we obtain
\begin{align}
    H'^{\dagger}&=GHG^{-1} =G(\eta^{-1}H^{\dagger}\eta)G^{-1} = G^{-1}H^{\dagger}G=H'
\end{align}
again using the $\eta$-Hermiticity of $H$.

\section{The $\eta$-adjoint of the unitary operator $U^{\ddagger}$}
Here, we illustrate the
$\eta-$adjoint of the $\eta$-unitary operator. We define $\tilde{\eta}=\eta\otimes I_B$ since $U$ operates on the composite Hilbert space of the system and the bath, and we need to exploit the system Hamiltonian $H$'s $\eta$-Hermiticity, $\eta^{-1}H^{\dagger}\eta=H$. Thus,
\begin{align}
    U^{\ddagger}&=\tilde{\eta}^{-1}U^{\dagger}\tilde{\eta} =\tilde{\eta}^{-1}e^{i\tilde{H}_U^\dagger t}\tilde{\eta}=\tilde{\eta}^{-1}\sum_{n=0}^\infty \frac{(i\tilde{H}_U^\dagger t)^n}{n!}\tilde{\eta} \nonumber \\
    &=\tilde{\eta}^{-1}\left(\sum_{n=0}^\infty \frac{(i t)^n}{n!}(\tilde{H}_U^\dagger.\tilde{H}_U^\dagger...) \right)\tilde{\eta} \nonumber \\
    &=\tilde{\eta}^{-1}\left(\sum_{n=0}^\infty \frac{(i t)^n}{n!}(\tilde{H}_U^\dagger \tilde{\eta}. \tilde{\eta}^{-1}\tilde{H}_U^\dagger...) \right)\tilde{\eta} \nonumber \\
    &=\left(\sum_{n=0}^\infty \frac{(i t)^n}{n!}(\tilde{\eta}^{-1}\tilde{H}_U^\dagger \tilde{\eta}. \tilde{\eta}^{-1}\tilde{H}_U^\dagger\tilde{\eta}...) \right) \nonumber \\
    &=\left(\sum_{n=0}^\infty \frac{(i t)^n}{n!}(\tilde{H}_U^\ddagger \tilde{H}_U^\ddagger...) \right)\nonumber \\ 
    &=\sum_{n=0}^\infty \frac{(i\tilde{H}_U^\ddagger t)^n}{n!} \nonumber \\ 
    &=\sum_{n=0}^\infty \frac{(i\tilde{H}_U t)^n}{n!} =e^{i\tilde{H}_U t}.
\end{align}
Here, the $\eta$-Hermiticity condition of $\tilde{H}_U$ is used \eqref{etahermiofComposite}.

\section{Derivation of generalized GKSL master equation for a $\pt$symmetric system}\label{sec_der_pt_gksl}

Here, we derive the GKSL master equation using Born-Markov and rotating wave approximations for a $\pt$symmetric system Hamiltonian coupled to a bosonic bath at thermal equilibrium. The Hamiltonian of the total system is given by
\begin{align}
    H_{\rm total} = H + H_B + H_{SB},
\end{align}
where $H$ is the $\pt$symmetric Hamiltonian, see Eq.~\eqref{nonhermiH}, $H_B$ is the bath Hamiltonian given by $H_B = \sum_k\omega_k a^\dagger_k a_k$, and $H_{SB} = \sum_k g_k \left(\sigma^+_\mathcal{G}\otimes a_k + \sigma^-_\mathcal{G} \otimes a^{\dagger}_k\right)$. In the interaction Hamiltonian, $\sigma^{+(-)}_\mathcal{G}$ denotes an excitation (de-excitation) of the $\pt$symmetric system, Eq.~\eqref{Biortho_sigmas}. Further, a Jaynes-Cummings type coupling between the system and the bath is considered, such that an excitation in the $\pt$symmetric system is accompanied by a photon decay and a de-excitation by a photon creation. As shown in Eq.~\eqref{etahermiofComposite}, the Hamiltonian $H_{\rm total}$ is $\eta$-Hermitian.
The initial state of the bath is taken to $\rho_B(0) = \exp(-\beta H_B)/{\rm Tr}\left[\exp(-\beta H_B)\right]$, where $\beta = 1/k_BT$ is the inverse temperature of the bath. To derive the master equation, we invoke the interaction picture, making use of the $\eta$-unitary operator $V(t) = e^{-i(H+H_B)t}$. In the interaction picture, $H_{SB}$ is transformed into 
\begin{align}
    \tilde H_{SB}(t) &= V^\ddagger (t) H_{SB} V(t) \nonumber \\
    &= e^{i(H+H_B)t}\left[\sum_kg_k\left(\sigma_{\mathcal{G}}^+a_k + \sigma_{\mathcal{G}}^-a_k^\dagger\right)\right]e^{-i(H+H_B)t},
\end{align}
which, using the spectral decomposition of $H$ in terms of $\ket{E_k}$ and $\ket{E_k^L}$  [Eq.~\eqref{SpectralDec}], becomes 
\begin{align}\label{eq_tilde_hsb}
    \tilde H_{SB}(t) = \sum_k g_k\left[\sigma_{\mathcal{G}}^+a_ke^{i(\Delta E - \omega_k)t} + \sigma_{\mathcal{G}}^-a_k^\dagger e^{-i(\Delta E - \omega_k)t}\right],
\end{align}
where $\Delta E = E_2 - E_1$.
Since $H_{\rm total }$ is $\eta$-Hermitian, it follows the generalized von Neumann equation, see Eq.~\eqref{eq_gen_von_neumann_eq},
\begin{align}
    \frac{d}{dt}\rho^{\rm tot}_{\mathcal{G}}(t) = -i\left[H_{\rm total}, \rho^{\rm tot}_\mathcal{G}(t)\right].
\end{align}
In the interaction picture, this generalized von Neumann equation becomes
\begin{align}\label{eq_interaction_von_Neumann}
    \frac{d}{dt}\tilde \rho^{\rm tot}_\mathcal{G}(t) = -\left[\tilde H_{SB}(t), \tilde \rho^{\rm tot}_\mathcal{G}(t)\right],
\end{align}
where the form of $\tilde H_{SB}(t)$ is given in Eq.~\eqref{eq_tilde_hsb}.
Integrating the above equation, we get 
\begin{align}
    \tilde \rho^{\rm tot}_\mathcal{G}(t) = \tilde \rho^{\rm tot}_\mathcal{G}(0) - i\int_0^tds\left[\tilde H_{SB}(s), \tilde \rho^{\rm tot}_{\mathcal{G}}(s)\right].
\end{align}
This solution is now substituted back into Eq.~\eqref{eq_interaction_von_Neumann}, and a partial trace $\tilde \rho_\mathcal{G}(t) = {\rm Tr}_B\left[\tilde \rho_\mathcal{G}^{\rm tot}(t)\right]$ is taken to obtain
\begin{align}
    \frac{d}{dt}\tilde \rho_\mathcal{G}(t) &= - i{\rm Tr}_B\left\{\left[\tilde H_{SB}(t), \tilde \rho_\mathcal{G}^{\rm tot}(0)\right]\right\} \nonumber \\
    &-{\rm Tr}_B\left\{\left[\tilde H_{SB}(t),\int_0^t ds \left[\tilde H_{SB}(s), \tilde \rho_\mathcal{G}^{\rm tot}(s)\right]\right]\right\}.
\end{align}
The first term in the above equation vanishes for an initial thermal state of the bath~\cite{lidar2020}, and thus we are left with
\begin{align}
    \frac{d}{dt}\tilde \rho_\mathcal{G}(t) &= -{\rm Tr}_B\left\{\left[\tilde H_{SB}(t),\int_0^t ds \left[\tilde H_{SB}(s), \tilde \rho_\mathcal{G}^{\rm tot}(s)\right]\right]\right\}.
    \label{eq_before_approx}
\end{align}

Now, we make our first approximation. For a sufficiently large bath compared to the system to which it is weakly coupled, it can be assumed that the bath remains stationary while the system undergoes non-trivial evolution~\cite{Breuer2007, lidar2020, Banerjee2018}. Hence, the state of the composite system at time $t$ can be given by
\begin{align}\label{eq_born_approx}
    \tilde \rho_\mathcal{G}^{\rm tot}(t) \approx \tilde \rho_\mathcal{G}(t) \otimes \rho_B,
\end{align}
where $\rho_B = \rho_B(0)$ is the time-independent thermal equilibrium state of the bath.
Substituting Eqs.~\eqref{eq_born_approx} in Eq.~\eqref{eq_before_approx}, we get 
\begin{align}\label{eq_after_Born_approx}
    &-\frac{d}{dt}\tilde \rho_\mathcal{G}(t) \nonumber \\&= {\rm Tr}_B\left\{\left[\tilde H_{SB}(t),\int_0^t ds \left[\tilde H_{SB}(s), \tilde \rho_\mathcal{G}(s)\otimes \rho_B\right]\right]\right\}.
\end{align}
On expanding the commutators, the right side of the above equation inside the integration $\int_0^t ds$ becomes
\begin{align}\label{eq_commutator_expansion}
    &{\rm Tr}_B\left\{\tilde H_{SB}(t)\tilde H_{SB}(s)\tilde \rho_\mathcal{G}(s)\rho_B - \tilde H_{SB}(t)\tilde \rho_\mathcal{G}(s)\rho_B\tilde H_{SB}(s)\right. \nonumber \\
    &\left.-\tilde H_{SB}(s)\tilde \rho_\mathcal{G}(s)\rho_B\tilde H_{SB}(t) + \tilde \rho_\mathcal{G}(s)\rho_B\tilde H_{SB}(s)\tilde H_{SB}(t)\right\}.
\end{align}
Before the next step, we simplify the Hamiltonian $\tilde H_{SB}(t)$, Eq.~\eqref{eq_tilde_hsb}, further as
\begin{align}\label{eq_tilde_hsb_redef}
    \tilde H_{SB}(t) = A(t)\otimes B(t) + A^\ddagger(t) \otimes B^\dagger (t),
\end{align}
where $A(t) = \sigma_\mathcal{G}^+e^{i\omega t}, A^\ddagger(t) = \sigma_\mathcal{G}^-e^{-i\omega  t}, B(t) = \sum_k g_k a_ke^{-i\omega_k t}$, and $B^\dagger(t) = \sum_k g_k a_k^\dagger e^{i\omega_k t}$, with $\omega = \Delta E$. We now substitute Eq.~\eqref{eq_tilde_hsb_redef} in Eq.~\eqref{eq_commutator_expansion} and simplify to rewrite Eq.~\eqref{eq_after_Born_approx} as 
\begin{align}\label{eq_commutator_expanded}
    -\frac{d}{dt}\tilde \rho_\mathcal{G}(t) = \int_0^t &ds \left\{\left[A(t), A^\ddagger(s)\tilde \rho_\mathcal{G}(s)\right] \langle B(t)B^\dagger(s)\rangle_B\right.\nonumber \\
    &\left.+ \left[A^\ddagger(t), A(s)\tilde \rho_\mathcal{G}(s)\right]\braket{B^\dagger(t)B(s)}_B\right.\nonumber \\
    &\left.+ \left[\tilde \rho_\mathcal{G}(s)A(s), A^\ddagger(t)\right]\braket{B(s)B^\dagger(t)}_B\right.\nonumber \\
    &\left.+\left[\tilde \rho_\mathcal{G}(s)A^\ddagger(s), A(t)\right]\braket{B^\dagger(s)B(t)}_B\right\},
\end{align}
where $\braket{(\cdot)}_B = {\rm Tr}_B[\rho_B(\cdot)]$. Let us now change the variables to $\tau = t - s$, such that $\int_0^t ds  = -\int_t^0d\tau = \int_0^td\tau$. Due to this transformation, we can simplify the correlation function $\braket{B(t)B^\dagger(s)}_B = \braket{B(t)B^\dagger(t-\tau)}_B$ as ${\rm Tr}_B\left[\rho_Be^{iH_B t}Be^{-iH_B t}e^{iH_B(t-\tau)}B^\dagger e^{-iH_B(t - \tau)}\right] = {\rm Tr}_B\left[\rho_B B(\tau)B^\dagger(0)\right] = \braket{B(\tau)B^\dagger}$. In a similar way, we can simplify other correlation functions as well and rewrite Eq.~\eqref{eq_commutator_expanded} as 
\begin{align}\label{eq_before_markov_approx}
    &-\frac{d}{dt}\tilde \rho_\mathcal{G}(t) \nonumber \\&= \int_0^t d\tau \left\{\left[A(t), A^\ddagger(t- \tau)\tilde \rho_\mathcal{G}(t- \tau)\right] \langle B(\tau)B^\dagger(0)\rangle_B\right.\nonumber \\
    &\left.+ \left[A^\ddagger(t), A(t- \tau)\tilde \rho_\mathcal{G}(t- \tau)\right]\braket{B^\dagger(\tau)B(0)}_B\right.\nonumber \\
    &\left.+ \left[\tilde \rho_\mathcal{G}(t- \tau)A(t- \tau), A^\ddagger(t)\right]\braket{B(0)B^\dagger(\tau)}_B\right.\nonumber \\
    &\left.+\left[\tilde \rho_\mathcal{G}(t- \tau)A^\ddagger(t- \tau), A(t)\right]\braket{B^\dagger(0)B(\tau)}_B\right\}.
\end{align}

In the above equation, the right side depends on the entire history of the system state as the $t-\tau$ in the state $\tilde \rho_\mathcal{G}(t-\tau)$ ranges from $t$ to 0 as $\tau$ increases from $0$ to $t$. This makes the equation time-nonlocal. To obtain a time-local differential equation for the system, we introduce the second approximation, the \textit{Markov} approximation. It states that the bath has a very short correlation time $\tau_B$. It is also assumed that the system bath coupling is weak $g_k\ll 1/\tau_B$ and the times $t$ are much larger than the bath correlation time $t\gg\tau_B$. Since the bath correlation function is zero for $\tau \gg \tau_B$ and since $t\gg \tau_B$, $\tilde \rho_\mathcal{G}(t-\tau)$ can be replaced by $\tilde \rho_\mathcal{G}(t)$. Under this approximation, Eq.~\eqref{eq_before_markov_approx} becomes
\begin{align}\label{eq_redfield_equation}
    -\frac{d}{dt}\tilde \rho_\mathcal{G}(t)= \int_0^t &d\tau \left\{\left[A(t), A^\ddagger(t- \tau)\tilde \rho_\mathcal{G}(t)\right] \langle B(\tau)B^\dagger(0)\rangle_B\right.\nonumber \\
    &\left.+ \left[A^\ddagger(t), A(t- \tau)\tilde \rho_\mathcal{G}(t)\right]\braket{B^\dagger(\tau)B(0)}_B\right.\nonumber \\
    &\left.+ \left[\tilde \rho_\mathcal{G}(t)A(t- \tau), A^\ddagger(t)\right]\braket{B(0)B^\dagger(\tau)}_B\right.\nonumber \\
    &\left.+\left[\tilde \rho_\mathcal{G}(t)A^\ddagger(t- \tau), A(t)\right]\braket{B^\dagger(0)B(\tau)}_B\right\},
\end{align}
which is the generalized \textit{Redfield} equation for the $\pt$symmetric system. Further, the upper limit of the integration in the above equation can be extended to infinity, that is, $\int_0^t d\tau \to \int_0^\infty d\tau$ under the Markov approximation. 

Now, using Eq.~\eqref{eq_tilde_hsb_redef} for $A(t)$ and $A^\ddagger(t)$ in  Eq.~\eqref{eq_redfield_equation}, we get
\begin{align}\label{eq_after_Markov_approximation}
    -\frac{d}{dt}\tilde \rho_\mathcal{G}(t) &= \Gamma_1(\omega)\left\{\sigma^+_\mathcal{G}\sigma_\mathcal{G}^- \tilde \rho_\mathcal{G}(t) - \sigma_\mathcal{G}^-\tilde \rho_\mathcal{G}(t)\sigma_\mathcal{G}^+ \right\}\nonumber \\
    &+\Gamma_2(\omega)\left\{  \sigma_\mathcal{G}^-\sigma_\mathcal{G}^+\tilde \rho_\mathcal{G}(t) - \sigma_\mathcal{G}^+\tilde\rho_\mathcal{G}(t)\sigma_\mathcal{G}^-\right\} \nonumber \\
    &+ \Gamma_1^*(\omega) \left\{\tilde \rho_\mathcal{G}(t)\sigma_\mathcal{G}^+\sigma^-_\mathcal{G} - \sigma^-_\mathcal{G}\tilde \rho_\mathcal{G}(t)\sigma^+_\mathcal{G}\right\} \nonumber \\
    &+ \Gamma_2^*(\omega) \left\{\tilde \rho_\mathcal{G}(t)\sigma^-_\mathcal{G}\sigma^+_\mathcal{G} - \sigma^+_\mathcal{G}\tilde \rho_\mathcal{G}(t)\sigma^-_\mathcal{G}\right\},
\end{align}
where $\Gamma_1(\omega) = \int_0^\infty d\tau e^{i\omega \tau}\braket{B(\tau)B^\dagger}_B$ and $\Gamma_2(\omega) = \int_0^\infty d\tau e^{-i\omega \tau}\braket{B^\dagger (\tau) B}_B$. Now, we can write $\Gamma_j(\omega) = \frac{1}{2}\gamma_j(\omega) + iS_j(\omega)$, and thus, $\gamma_j(\omega) = \Gamma_j(\omega) + \Gamma_j^*(\omega)$ and $S_j(\omega) = \frac{-i}{2}\left[\Gamma_j(\omega) - \Gamma_j^*(\omega)\right]$ for $j = \{1, 2\}$. The function $\gamma_j(\omega)$ and $S_j(\omega)$ are well defined in the literature~\cite{Breuer2007, lidar2020, Banerjee2018}. For a two-level system, we have $\gamma_1(\omega) = \gamma_0[1 + N(\omega)]$ and $\gamma_2(\omega) = \gamma_0 N(\omega)$, where $\gamma_0$ is the spontaneous emission rate and $N(\omega) = 1/\left(e^{\beta\omega} - 1\right)$ is the Planck distribution. Using these $\gamma_j(\omega)$ in Eq.~\eqref{eq_after_Markov_approximation} and defining the Lamb shift term $H_{LS} = S_1(\omega)\sigma^+_\mathcal{G}\sigma^-_\mathcal{G} + S_2(\omega)\sigma^-_\mathcal{G}\sigma^+_\mathcal{G}$, we can write the generalized GKSL master equation for a $\pt$symmetric open quantum system in the interaction picture as 
\begin{align}
    \frac{d}{dt}\tilde \rho_\mathcal{G}(t) &= -i\left[H_{LS}, \tilde \rho_\mathcal{G}(t)\right] \nonumber \\
    &+ \gamma_0[1 + N(\omega)]\left[\sigma^-_\mathcal{G}\tilde \rho_\mathcal{G}(t)\sigma^+_\mathcal{G} - \frac{1}{2}\left\{\sigma^+_\mathcal{G}\sigma^-_\mathcal{G}, \tilde \rho_\mathcal{G}(t)\right\}\right] \nonumber \\
    &+ \gamma_0 N(\omega)\left[\sigma^+_\mathcal{G}\tilde \rho_\mathcal{G}(t)\sigma^-_\mathcal{G} - \frac{1}{2}\left\{\sigma^-_\mathcal{G}\sigma^+_\mathcal{G}, \tilde \rho_\mathcal{G}(t)\right\}\right].
\end{align}
We now go back to the Schr\"{o}dinger picture by applying the transformation $\rho_\mathcal{G}(t) = V(t)\tilde \rho_\mathcal{G}(t)V^\ddagger(t)$, where $V(t) = e^{-i Ht}$. In this picture, the generalized GKSL master equation is
\begin{align}
    \frac{d}{dt} \rho_\mathcal{G}(t) &= -i\left[H + H_{LS},  \rho_\mathcal{G}(t)\right] \nonumber \\
    &+ \gamma_0[1 + N(\omega)]\left[\sigma^-_\mathcal{G} \rho_\mathcal{G}(t)\sigma^+_\mathcal{G} - \frac{1}{2}\left\{\sigma^+_\mathcal{G}\sigma^-_\mathcal{G},  \rho_\mathcal{G}(t)\right\}\right] \nonumber \\
    &+ \gamma_0 N(\omega)\left[\sigma^+_\mathcal{G} \rho_\mathcal{G}(t)\sigma^-_\mathcal{G} - \frac{1}{2}\left\{\sigma^-_\mathcal{G}\sigma^+_\mathcal{G},  \rho_\mathcal{G}(t)\right\}\right].
\end{align}
Interestingly, the structure of this generalized $\pt$symmetric GKSL master equation is similar to the one for a dissipative two-level Hermitian system, with $\rho_\mathcal{G}(t)$, $\sigma_\mathcal{G}^\pm(t)$, and $\pt$symmetric Hamiltonian $H$ taking places of the density matrix, jump operators, and system Hamiltonian in the regular GKSL master equation. 
All the decoherence rates of this master equation are constant and positive, making the master equation appropriate for a semigroup evolution, which denotes a completely positive and trace preserving map.

It is noteworthy that the final form and the derivation of the above master equation may appear like that of the standard GKSL master equation, but they differ in a fundamental sense as they involve $\eta$-Hermitian system operators having $\pt$symmetry. Despite the projectors of such system operators having a different structure than that of a usual system operator for a Hermitian Hamiltonian evolution (having influences of the $\pt$symmetric Hamiltonian parameters), they conform to the structure of the usual GKSL master equation.

\section{Redefining Pauli operators in the bi-orthonormal basis}
In Eqs.~(\ref{Biortho_sigmas}) and ~(\ref{biorthoSigmas}), we see how the bi-orthonormal basis vectors build the suitable Pauli $\sigma_{\mathcal{G}}$ operators.
It should be noted that $\sigma_{\mathcal{G}}$'s are not linked with the usual Pauli $\sigma$ matrices with relations like $\sigma_{\mathcal{G}}=\sigma\eta$, i.e., the bi-orthonormal Pauli matrices are not obtained by operating the $\eta$ in Eq.~(\ref{eta_structure}) by means of a matrix product to the $\sigma$ matrices based on the computational basis.
The function of $\eta$ is to modify the projectors of any two-level operators in the bi-orthonormal space by virtue of equation (\ref{eta_function}). Let us take $\sigma_{\mathcal{G}}^+$ as an example
\begin{align}
    \sigma_{\mathcal{G}}^+&=\ket{E_2^R}\bra{E_1^L} =\ket{E_2^R}\bra{E_1^L}\eta.
\end{align}
The $\ket{E_2}\bra{E_1}$ is obtained using the right eigenvectors listed in Eq.~(\ref{righteigs}). The operator $\sigma^+$, on the other hand, is simply
\begin{align}
    \sigma^+&=\ket{0}\bra{1} =\begin{pmatrix}
                1\\
                0
            \end{pmatrix}\begin{pmatrix}
                0 & 1
            \end{pmatrix} =\begin{pmatrix}
                0&1 \\
                0&0
            \end{pmatrix}.
\end{align}
Therefore, it is evident from the above that 
\begin{equation}
    \sigma_{\mathcal{G}}^+\neq\sigma^+\eta.
\end{equation}

The bi-orthonormal Pauli operators follow the commutation relation $[\sigma_{\mathcal{G}}^i, \sigma_{\mathcal{G}}^j] = 2i \, \varepsilon^{ijk} \, \sigma_{\mathcal{G}}^k$.
As an example, we calculate $[\sigma^x_{\mathcal{G}},\sigma^y_{\mathcal{G}}]$ to show the validity of the above commutation relation.
We have
\begin{align}
    \sigma^x_{\mathcal{G}}\sigma^y_{\mathcal{G}}&=(\ket{E_1^R}\bra{E_2^L}+\ket{E_2^R}\bra{E_1^L} )(i\ket{E_1^R}\bra{E_2^L}-i\ket{E_2^R}\bra{E_1^L} ) \nonumber \\
    &=-i\ket{E_1^R}\bra{E_1^L}+i\ket{E_2^R}\bra{E_2^L},
\end{align}
and
\begin{align}
    \sigma^y_{\mathcal{G}}\sigma^x_{\mathcal{G}}&=(i\ket{E_1^R}\bra{E_2^L}-i\ket{E_2^R}\bra{E_1^L} )(\ket{E_1^R}\bra{E_2^L}+\ket{E_2^R}\bra{E_1^L} ) \nonumber \\
    &=i\ket{E_1^R}\bra{E_1^L}-i\ket{E_2^R}\bra{E_2^L}.
\end{align}
Subtracting them, we get
\begin{align}
    [\sigma^x_{\mathcal{G}},\sigma^y_{\mathcal{G}}]&=-2i\ket{E_1^R}\bra{E_1^L}+2i\ket{E_2^R}\bra{E_2^L} \nonumber \\
    &=2i\sigma^z_{\mathcal{G}},
\end{align}
which verifies the above commutation relation.

\bibliographystyle{apsrev4-1}
\bibliography{references}
\end{document}